\documentclass[aps,prl,twocolumn,longbibliography,10pt]{revtex4-2}

\usepackage{amsmath}
\usepackage{amssymb}
\usepackage{graphicx}
\usepackage[colorlinks=true,
urlcolor=cyan]{hyperref}
\usepackage{bm}
\usepackage{subfigure}
\usepackage{xcolor}
\usepackage{enumitem}
\usepackage{amsthm}

\newcommand{\N}{\mathbb{N}}
\newcommand{\Z}{\mathbb{Z}}
\newcommand{\C}{\mathbb{C}}
\newcommand{\EE}{\mathcal{E}}
\newcommand{\NN}{\mathcal{N}}
\newcommand{\Tr}{\mathrm{Tr}}
\newcommand{\TT}{\mathcal{T}}
\newcommand{\VV}{\mathcal{V}}

\definecolor{darkblue}{rgb}{0.0,0.28,0.85}

\makeatletter
\newenvironment{theorem*}{%
  \begin{trivlist}%
    \item[\hskip\labelsep\bfseries Theorem\@addpunct{.}]\itshape
}{%
  \end{trivlist}%
}
\makeatother

\makeatletter
\newenvironment{lemma*}{%
  \begin{trivlist}%
    \item[\hskip\labelsep\bfseries Lemma\@addpunct{.}]\itshape
}{%
  \end{trivlist}%
}
\makeatother

\begin{document}

\title{Robust Mixed-State Cluster States and Spurious Topological Entanglement Negativity}

\author{Seunghun Lee}
\affiliation{Department of Physics, Korea Advanced Institute of Science and Technology, Daejeon 34141, Republic of Korea}
\author{Eun-Gook Moon}
\thanks{egmoon@kaist.ac.kr}
\affiliation{Department of Physics, Korea Advanced Institute of Science and Technology, Daejeon 34141, Republic of Korea}

\begin{abstract}
    We investigate 1D and 2D cluster states under local decoherence to assess the robustness of their mixed-state subsystem symmetry-protected topological (SSPT) order. By exactly computing fidelity correlators via dimensional reduction of effective statistical mechanics models, we pinpoint the critical error rate for strong-to-weak spontaneous breaking of strong subsystem symmetry. Without resorting to the replica trick, we demonstrate that mixed-state SSPT order remains remarkably robust up to the maximal decoherence rate when noise respects strong subsystem symmetry. Furthermore, we propose that the mixed-state SSPT order can be detected by a constant correction to the area-law scaling of entanglement negativity, termed spurious topological entanglement negativity. This also highlights that topological entanglement negativity, a widely used diagnostic for mixed-state topological order, is generally not invariant under finite-depth quantum channels. 
\end{abstract}

\maketitle

%================================%
\textit{Introduction}---Symmetry and topology have been two conceptual pillars for understanding quantum phases of matter~\cite{wen2004quantum,sachdev2023quantum}, broadly categorizing ground states of local gapped Hamiltonians into intrinsic topological order, symmetry-protected topological (SPT) order, and symmetry-broken order. Recently, driven by experimental progress in quantum platforms, the study of quantum phases has been extended to mixed states under local decoherence~\cite{coser2019classification,de2022symmetry,lee2022symmetry,fan2024diagnostics,bao2023mixed,sang2024mixed,sang2024stability,zou2023channeling,guo2024two,lu2023mixed,ma2024symmetry,sun2024holographic,ellison2024towards,sohal2024a,min2024mixed,chen2023symmetry,chen2024separability,wang2307intrinsic,zhang2024quantum,you2024intrinsic,lu2024bilayer,su2024higher,sala2024stability,shah2024instability,kim2024persistent,lee2024mixed,li2024replica,ma2023average,xue2024tensor,guo2024locally,ding2024boundary,wang2025analog,lee2023quantum,lessa2024strong,sala2024spontaneous,ma2023topological,kuno2024strong,zhang2024strong,chen2024strong,gu2024spontaneous,huang2024hydrodynamics,weinstein2024efficient,liu2024diagnosing,kim2024error,guo2025quantum}, which naturally emerge on current noisy quantum platforms. This paradigm shift has opened new avenues for understanding the interplay between symmetry and topology in open quantum systems.

For mixed states, a salient difference from the pure-state case is the two distinct notions of symmetry~\cite{buvca2012a}: Given a symmetry transformation $U$, a mixed state $\rho$ exhibits \emph{strong symmetry} when $U\rho = e^{i\theta} \rho$, i.e., $\rho$ carries a well-defined symmetry charge. In contrast, $\rho$ exhibits \emph{weak symmetry} when $U\rho U^\dagger = \rho$, meaning that $\rho$ is an ensemble of states with varying symmetry charges. In the thermodynamic limit, the strong symmetry can spontaneously break into weak symmetry---a phenomenon unique to mixed states, known as \emph{strong-to-weak spontaneous symmetry breaking} (SWSSB)~\cite{lee2023quantum,lessa2024strong,sala2024spontaneous,ma2023topological,kuno2024strong,zhang2024strong,chen2024strong,gu2024spontaneous,huang2024hydrodynamics,weinstein2024efficient,liu2024diagnosing,kim2024error,guo2025quantum}. Mixed-state phase transitions induced by local decoherence can be understood in terms of SWSSB of the symmetry associated with the quantum state~\cite{lee2023quantum,lessa2024strong,sala2024spontaneous} and can be detected by information-theoretic quantities such as the fidelity correlator~\cite{lessa2024strong}. These developments raise questions about the robustness of interesting quantum states against SWSSB and how symmetry affects the entanglement structure of decohered mixed states.

In this work, we consider 1D and 2D cluster states subjected to local decoherence. Cluster states are crucial resource states for various quantum information processing tasks, such as measurement-based quantum computation (MBQC)~\cite{raussendorf2001a,raussendorf2003measurement,else2012symmetry1,stephen2017computational,raussendorf2017symmetry,doherty2009identifying,else2012symmetry2,raussendorf2019computationally} and the preparation of long-range entangled states~\cite{verresen2021efficiently,lu2022measurement,lee2022decoding,zhu2023nishimoris,tantivasadakarn2023hierarchy,tantivasadakarn2023shortest}. Their utility stems from their nontrivial subsystem symmetry-protected topological (SSPT) order~\cite{you2018subsystem,devakul2018classification}, which is protected by symmetries acting on rigid, line-like subsystems. We analyze the fidelity correlator of cluster states to assess the robustness of their mixed-state SSPT order against SWSSB. Through mapping to statistical mechanics (stat-mech) models, we demonstrate the remarkable robustness of mixed-state SSPT order up to the maximal error rate $p = 1/2$ under a \emph{general} local Pauli noise that respects strong subsystem symmetry. We also provide numerical evidence that this robustness extends to local non-Pauli noise. This robustness can be understood from the dimensional reduction applied to effective stat-mech models, which inherit subsystem symmetries from the decohered mixed states. Notably, our arguments do not rely on the replica trick and are thus free from the subtleties associated with the doubled Hilbert space formalism~\cite{lessa2024strong}.

Furthermore, we investigate how strong subsystem symmetry affects the mixed-state entanglement of decohered density matrices, focusing on the entanglement (logarithmic) negativity $\EE_R$~\cite{vidal2002computable,plenio2005logarithmic}. Similar to the entanglement entropy for ground states of local gapped Hamiltonians, the entanglement negativity of locally decohered states is expected to scale as~\cite{grover2011entanglement,lu2020detecting}
\begin{align} \label{TEN}
    \EE_R = \alpha' |\partial R| - \EE_{\text{topo}} + \cdots,
\end{align}
where $|\partial R|$ is the perimeter of a region $R$, and the constant correction $\EE_{\text{topo}}$, termed the topological entanglement negativity (TEN), has been used to diagnose mixed-state topological order~\cite{lu2020detecting,fan2024diagnostics,lu2023characterizing,lu2024disentangling,lu2020structure,wu2020entanglement}. For 1D and 2D cluster states, we demonstrate the existence of a spurious topological contribution to the area-law scaling, which we dub \emph{spurious TEN} $\EE_{\text{sp}}$, despite the short-range entanglement of decohered density matrices. We further show that spurious TEN generally arises when the decohered density matrix retains strong subsystem symmetry. This result implies that, in most general cases, TEN is not invariant under a finite-depth local quantum channel. The discovered spurious TEN is an information-theoretic quantity from the bulk that captures the mixed anomaly between strong subsystem symmetries and is reminiscent of the spurious topological entanglement entropy (TEE) observed in pure states~\cite{cano2015interactions,santos2018symmetry,kato2020toy,zou2016spurious,stephen2019detecting,devakul2018classification,williamson2019spurious}.

\begin{figure}[t]
    \includegraphics[width=\columnwidth]{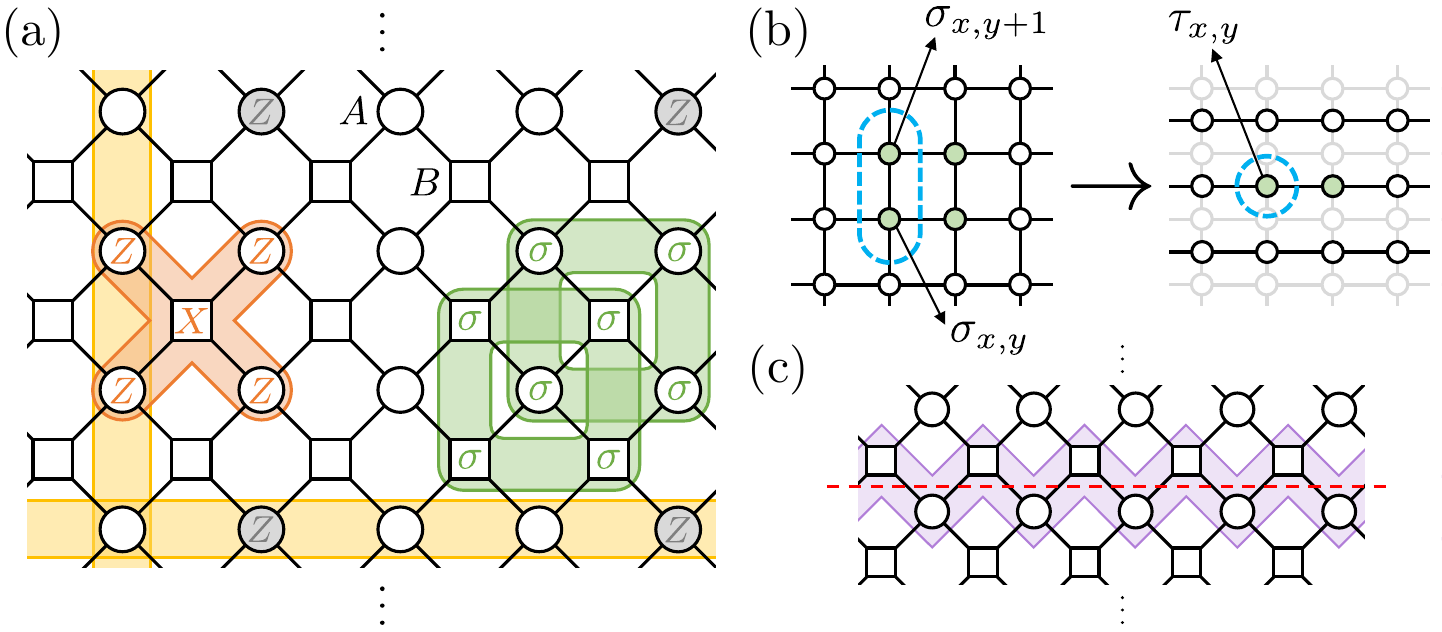}
    \caption{(a) Cluster state on a 2D square lattice with an infinite cylindrical geometry (periodic along the horizontal direction). Circles (squares) denote the $A$ ($B$) sublattice. The orange shape represents the stabilizer $K_j = X_j \prod_{i\in\partial j} Z_i$, yellow lines show the supports of subsystem symmetry generators, and gray circles mark four charged $Z$-operators for the fidelity correlator $F_Z^{\text{2D}}$ [see Eq.~\eqref{FC_subsystem}]. Green shapes indicate the interaction $\prod_{i\in \square_{A/B}} \sigma_i$ in the emergent plaquette Ising model. (b) Mapping of the 2D plaquette Ising model to stacks of 1D Ising models, with the blue dashed curves showing the redefinition of Ising variables ($\sigma \rightarrow \tau$). (c) Bipartition for entanglement negativity, where the purple shape depicts the 1D cluster state after locally disentangling other qubits.}
    \label{fig:2DCluster}
\end{figure}

%================================%
\textit{Fidelity Correlator: Noisy 1D Cluster State}---We begin with 1D cluster state~\cite{briegel2001persistent} on a periodic chain of $2N$ qubits, with Pauli operators $X_j$ and $Z_j$ at site $j$. The 1D cluster state $| \psi_0 \rangle$ is the ground state of the Hamiltonian $H_{\text{1D}} = -\sum_{j = 1}^{2N} K_j$, where $K_j = Z_{j-1} X_j Z_{j+1}$. Let $A$ ($B$) denote the sublattice of odd (even) sites. The state $|\psi_0 \rangle$ has a $\Z_2 \times \Z_2$ SPT order protected by subsystem symmetries $G_A = \prod_{j=1}^N X_{2j-1}$ and $G_B = \prod_{j=1}^N X_{2j}$~\cite{son2012topological,chen2014symmetry}. (We refer to these onsite symmetries as ``subsystem symmetries'' because line-like symmetries of the same form protect the SSPT order of the 2D cluster state on a square lattice; see Fig.~\ref{fig:2DCluster}.) Let $\rho_0 = | \psi_0 \rangle \langle \psi_0|$ represent the pure cluster state. Subjecting each qubit to a quantum channel $\NN_j^P [\rho] = (1-p) \rho + p P_j \rho P_j$ ($P = X, Z$), with $p$ the error rate, the resulting decohered mixed state is given by $\rho_P = \prod_{j=1}^{2N} \NN_j^P [\rho_0]$.

Notice that $\rho_X$ exhibits a strong symmetry, satisfying $G_i \rho_X = \rho_X$, whereas $\rho_Z$ only possesses a weak symmetry, i.e., $G_i \rho_Z G_i^\dagger = \rho_Z$ for $p > 0$. A mixed state $\rho$ with strong symmetry can undergo SWSSB if its strong symmetry is spontaneously broken to weak symmetry in the thermodynamic limit~\cite{lee2023quantum,lessa2024strong,sala2024spontaneous,one}. While conventional spontaneous symmetry breaking is defined by the long-range order in the correlation function $\Tr [\rho O_x O_y^\dagger]$ of local charged operators $O_{x,y}$, SWSSB is identified by the long-range order in the fidelity correlator~\cite{lessa2024strong} 
\begin{align} \label{FC}
    F_O (x,y) = F(\rho, O_x O_y^\dagger \rho O_x^\dagger O_y),
\end{align}
where $F(\rho, \sigma) = \Tr \left[\sqrt{\rho^{1/2} \sigma \rho^{1/2}} \right]$ is the fidelity between the two states $\rho$ and $\sigma$. Namely, a mixed state $\rho$ with strong symmetry has SWSSB when $F_O(x,y)$ approaches a finite constant as $|x-y| \rightarrow \infty$, while the conventional correlation function vanishes. The robustness of the strong symmetries $G_{A,B}$---and thereby the mixed-state SPT order---against SWSSB can be assessed through the fidelity correlator.

To compute the fidelity correlator for $\rho_X$, note that $| \psi_0 \rangle$ is a stabilizer state satisfying $K_j |\psi_0 \rangle = | \psi_0 \rangle$, with its density matrix given by $\rho_0 \propto \sum_{\{a\}} \prod_{j=1}^{2N} K_j^{a_j}$, where $a_j \in \{0,1\}$ indicates the presence or absence of the stabilizer $K_j$. Under conjugation by $X_i$, $\prod_{j=1}^{2N} K_j^{a_j}$ flips (maintains) its sign when $\sigma_{i-1} \sigma_{i+1} = -1$ ($+1$), where $\sigma_j = 1 - 2a_j \in \{\pm 1\}$. From this, the spectrum of the decohered density matrix can be expressed as
\begin{align} \label{rhoXK}
    \rho_X (\{K\}) \propto \sum_{\{a\}} \prod_{j=1}^{2N} K_j^{a_j} (1-2p)^{\sum_{i=1}^{2N} \frac{1 - \sigma_i \sigma_{i+2}}{2}}.
\end{align}
Defining $\beta = -\frac 12 \log (1 - 2p)$ and using $K_j^{a_j} = \sigma_j^{(1 -K_j)/2}$, Eq.~\eqref{rhoXK} becomes $\rho_X (\{K\}) \propto \langle \sigma_S \rangle_\beta$, where $\sigma_S \equiv \prod_{i\in S} \sigma_i$ is the product of variables $\sigma_i$ in $S = \{j: K_j = -1\}$, and $\langle \cdot \rangle_\beta = Z_\beta^{-1} \sum_{\{\sigma\}} (\cdot) e^{\beta \sum_{i=1}^{2N} \sigma_i \sigma_{i+2}}$ denotes the expectation value in two independent Ising chains at inverse temperature $\beta$ (with $Z_\beta$ being the corresponding partition function). In the context of quantum error correction, the subset $S$ corresponds to the error syndrome, which must contain an even number of elements per sublattice due to the subsystem symmetry of $|\psi_0\rangle$. For $O_{x,y} = Z_{x,y}$ with $x, y \in A$, it can be similarly shown that the spectrum of $O_x O_y^\dagger \rho_X O_x^\dagger O_y$ is proportional to $\langle \sigma_{S \triangle \{x, y\}} \rangle_\beta$, where $\triangle$ denotes the symmetric difference~\cite{two}.

\begin{figure}[t]
    \includegraphics[width=\columnwidth]{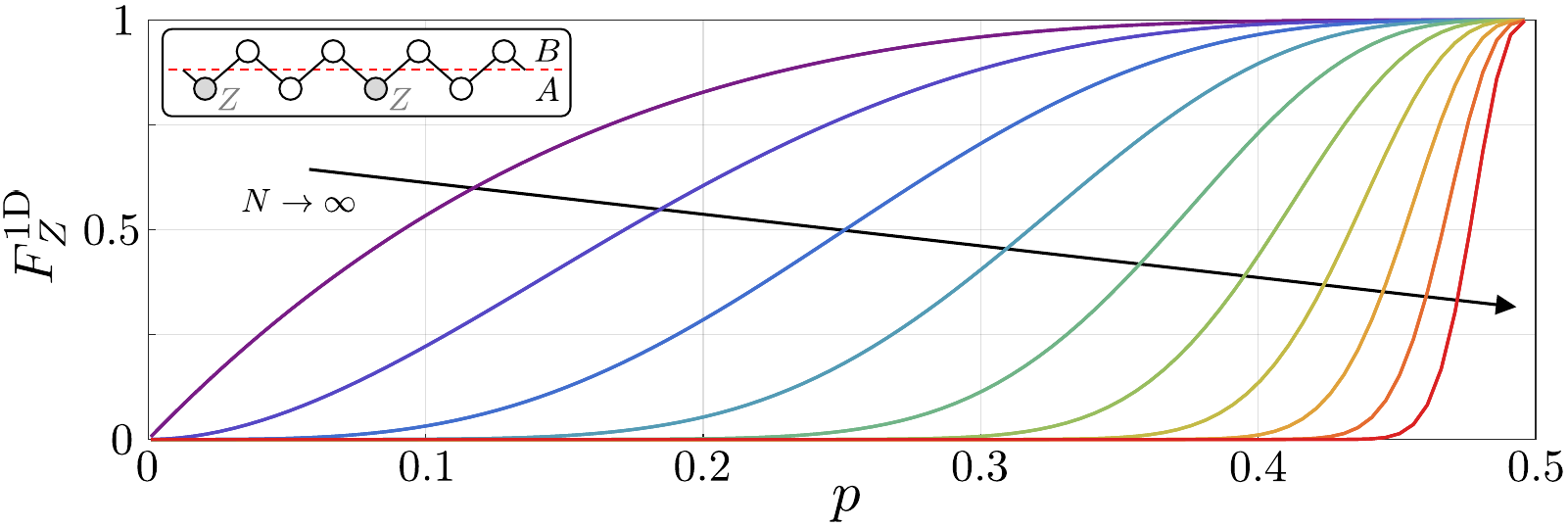}
    \caption{Fidelity correlator $F_Z^{\mathrm{1D}}(x,y)$ with $|x-y| = N$ for a 1D cluster state of length $2N$ under $X$-noise [see Eq.~\eqref{1D_FC}]. Each curve corresponds to $N = 2^n$, with $n$ ranging from 1 (purple) to 10 (red). The inset shows the locations of the charged operators $Z_{x,y}$ for the case $N = 4$.}
    \label{fig:1DCluster_FC}
\end{figure}

Substituting these two expressions into Eq.~\eqref{FC} gives
\begin{align} \label{1D_FC}
    \! F_Z^{\mathrm{1D}} (x,y) &= \frac{\sum_S \left[ \langle \sigma_S \rangle_\beta \langle \sigma_{S \triangle \{x,y\}} \rangle_\beta \right]^{1/2}}{\sum_S \langle \sigma_S \rangle_\beta}, 
\end{align}
where $S$ runs over all subsystem symmetric subsets of the chain~\cite{three}. Therefore, $F_Z^{\mathrm{1D}}(x,y)$ maps to an ``overlap'' between two distributions proportional to multi-spin correlators of the 1D Ising model, where the subsets differ only by $\{x,y\}$. In the Supplementary Material (SM)~\cite{SM}, we analytically compute Eq.~\eqref{1D_FC} for finite $N$ and confirm the exponential decay of $F_Z^{\mathrm{1D}} (x,y)$ to zero for $p < 1/2$~\cite{four}. In contrast, $F_Z^{\mathrm{1D}} (x,y) = 1$ at $p = 1/2$, indicating that the 1D cluster state under $X$-noise undergoes SWSSB only at $p = 1/2$. See Fig.~\ref{fig:1DCluster_FC} for $F_Z^{\mathrm{1D}}(x,y)$ with $|x-y| = N$ at various system sizes $2N$. 

This robustness of the mixed-state 1D SPT order under $X$-noise extends to \emph{any} local incoherent Pauli noises $\NN_j^P [\rho] = (1-p) \rho + p P_j \rho P_j$ that respect strong subsystem symmetry. (Here, $P_j$ may act on a finite number of qubits near site $j$.) In SM~\cite{SM}, we show that, with an appropriate choice of charged operators $O_{x,y}$, the fidelity correlator of the 1D cluster state decohered under any such noise maps to a stat-mech expression analogous to Eq.~\eqref{1D_FC}, which decays exponentially for $p < 1/2$. Moreover, we present numerical evidence in SM~\cite{SM} showing that this robustness extends to strongly symmetric local non-Pauli noise. We therefore expect the mixed-state SPT order of the 1D cluster state to remain robust under general local decoherence that respects strong subsystem symmetry. We note that our stability analysis from the viewpoint of SWSSB is consistent with the tensor network analysis of Ref.~\cite{xue2024tensor}.

%================================%
\textit{Fidelity Correlator: Noisy 2D Cluster State}---Next, we consider a cluster state on a 2D square lattice defined on an infinite cylindrical geometry [see Fig.~\ref{fig:2DCluster}(a)], which is a ground state of the Hamiltonian $H_{\text{2D}} = -\sum_j K_j$ with $K_j = X_j \prod_{i\in \partial j} Z_i$, where $i\in\partial j$ denotes the four vertices adjacent to vertex $j$. The 2D cluster state hosts a $\Z_2^{\text{sub}}$ SSPT order~\cite{doherty2009identifying,else2012symmetry2,raussendorf2019computationally,devakul2018classification,you2018subsystem}, with extensively many symmetry generators $\prod_{j\in \text{diag}} X_j$ along each diagonal lines [yellow lines in Fig.~\ref{fig:2DCluster}(a)]. Let's consider the decohered mixed state $\rho_X$ under the $X$-noise, which respects strong $\Z_2^{\text{sub}}$ subsystem symmetry.

We can detect SWSSB of the $\Z_2^{\text{sub}}$ subsystem symmetry using the following fidelity correlator~\cite{sala2024spontaneous}:
\begin{align} \label{FC_subsystem}
    F_Z^{\text{2D}} (w,h) = F \!\left( \rho, \prod_{i\in \square_{wh}} Z_i \rho \prod_{i\in \square_{wh}} Z_i \right),
\end{align}
where $\square_{wh}$ is four corners of a square with width $w$ and height $h$ [gray circles in Fig.~\ref{fig:2DCluster}(a)]. We can rewrite Eq.~\eqref{FC_subsystem} for $\rho_X$ in terms of a stat-mech model, similar to the 1D case. Since $\prod_j K_j^{a_j}$ ($a_j \in\{0,1\}$) flips (maintains) its sign under conjugation by $X_j$ when $\prod_{i\in \partial j} \sigma_i = -1$ ($+1$), we have
\begin{align} \label{2D_rhoX}
    \rho_X \propto \sum_{\{a\}} \prod_j K_j^{a_j} e^{\beta \left(\sum_{\square_A} \prod_{i \in \square_A} \sigma_i + \sum_{\square_B} \prod_{i \in \square_B} \sigma_i \right)},
\end{align}
where $\beta = -\frac 12 \log (1-2p)$, $\sigma_j = 1 - 2a_j \in \{\pm 1\}$, and $\square_{A/B}$ are elementary plaquettes in the $A$/$B$ sublattice [green shapes in Fig.~\ref{fig:2DCluster}(a)]. Thus, $F_Z^{\text{2D}}$ follows an expression similar to Eq.~\eqref{1D_FC}, with $\{x,y\}$ replaced by $\square_{wh}$ and expectation values taken for two independent copies of the 2D plaquette Ising models (PIMs)~\cite{bathas1995two}. Note that the PIM inherits line-like subsystem symmetries from $\rho_X$. 

Now, consider a single copy of the PIM on a cylinder of height $L$. Defining new Ising variables as $\tau_{x,1} = \sigma_{x,1}$ and $\tau_{x,y} = \sigma_{x,y-1} \sigma_{x,y}$ for $2\leq y\leq L$ along each column [see Fig.~\ref{fig:2DCluster}(b)], the four-body interaction $\prod_{i \in \square} \sigma_i$ maps to two-body interaction $\tau_{x,y} \tau_{x+1,y}$, reducing the PIM to a stack of 1D Ising models with periodic boundary conditions in the large $L$ limit~\cite{mueller2017exact,five}. From this dimensional reduction from 2D to 1D, we have~\cite{SM},
\begin{align} \label{DimRed}
    F_Z^{\text{2D}} (w, h) \underset{L \rightarrow \infty}{\longrightarrow} [F_Z^{\text{1D}} (h)]^2
\end{align}
 for $p < 0.5$, where $F_Z^{\text{1D}} (h)$ is the fidelity correlator Eq.~\eqref{1D_FC} for the $X$-decohered 1D cluster state  with $|x-y| 
= h$. Since $F_Z^{\text{1D}} (h)$ decays exponentially with $h$, $F_Z^{\text{2D}} (h,w)$ also decays exponentially with $h$ in the thermodynamic limit. In other words, the mixed-state SSPT order of the 2D cluster state remains robust against SWSSB up to the maximal error rate $p = 1/2$ under $X$-noise. 

This robustness of the mixed-state SSPT order in the 2D cluster state extends to \emph{any} local incoherent Pauli noise that preserves strong subsystem symmetry. We show in SM~\cite{SM} that the associated stat-mech models for the fidelity correlator exhibit line-like subsystem symmetries that originate from the strong subsystem symmetry of the decohered cluster state. These symmetries enable a redefinition of spin variables that reduces the model to a stack of 1D Ising models. Consequently, as Eq.~\eqref{DimRed}, the fidelity correlator factorizes into a product of those for decohered 1D cluster states. Therefore, it generally exhibits exponential decay for $p < 1/2$, establishing the stability of the mixed-state SSPT order up to the maximal error rate $p = 1/2$. One can apply the same analysis to a 2D cluster state on a triangular lattice to yield the same conclusion.

%================================%
\textit{Spurious Topological Entanglement Negativity}---We now investigate the entanglement negativity of decohered 1D cluster states. The entanglement negativity~\cite{vidal2002computable,plenio2005logarithmic} is a mixed-state entanglement measure defined as $\EE_R = \log \| \rho^{T_R} \|_1$, where $T_R$ denotes a partial transpose on region $R$ and $\| \cdot \|_1$ is a trace norm. Let's start with the 1D cluster state $\rho_X$ under $X$ noise. To compute the spectrum of the partially transposed density matrix $(\rho_X)^{T_A} \propto \sum_{\{a\}} \!\left[ \prod_{j=1}^{2N} K_j^{a_j} \right]^{T_A} e^{\beta\sum_{i=1}^{2N} \sigma_i \sigma_{i+2}}$, we make the following observation: when two adjacent $K_j$ stabilizers are present simultaneously, the partial transpose on the sublattice $A$ gives a factor of $-1$, corresponding to a Pauli-$Y$ operator on $A$~\cite{lu2020detecting,lu2023characterizing,lu2024disentangling}. This leads to $\left[ \prod_{j=1}^{2N} K_j^{a_j} \right]^{T_A} = \prod_{j=1}^{2N} K_j^{a_j} (-1)^{\sum_{i=1}^{2N} a_i a_{i+1}}$ and hence
\begin{align} \label{rhoDX_TA}
    (\rho_X)^{T_A} \propto \sum_{\{a\}} \prod_{j=1}^{2N} K_j^{a_j} (-1)^{\sum_{i=1}^{2N} a_i a_{i+1}} e^{\beta\sum_{i=1}^{2N} \sigma_i \sigma_{i+2}}.
\end{align}
Using the relations $K_j^{a_j} = \sigma_j^{(1 -K_j)/2}$ and $(-1)^{a_i a_{i+1}} = \exp \left[\frac{i\pi}{4} (1 - \sigma_i - \sigma_{i+1} + \sigma_i \sigma_{i+1}) \right]$, the trace norm of the partial-transposed density matrix can be recast as
\begin{align} \label{1D_TraceNorm}
    \| (\rho_X)^{T_A} \|_1 = \frac{1}{C_X} \sum_S \left| \left\langle \prod_{i\in S} \sigma_i \right\rangle_{\!\beta,X} \right|,
\end{align}
where $S$ runs over all subsets of the chain and $C_X$ is a constant that depends on $\beta$ and $N$~\cite{SM}. Here, $\langle \cdot \rangle_{\beta,X}$ denotes the expectation value in the non-Hermitian 1D stat-mech model defined by $H_{\beta,X} = \sum_{i=1}^{2N} \left( \frac{i\pi}{4} \sigma_i \sigma_{i+1} + \beta \sigma_i \sigma_{i+2} - \frac{i\pi}{2} \sigma_i \right)$. Similarly, the trace norm of $\rho_Z$ can also be expressed in terms of a non-Hermitian stat-mech model~\cite{SM}.

\begin{figure}[t]
    \includegraphics[width=\columnwidth]{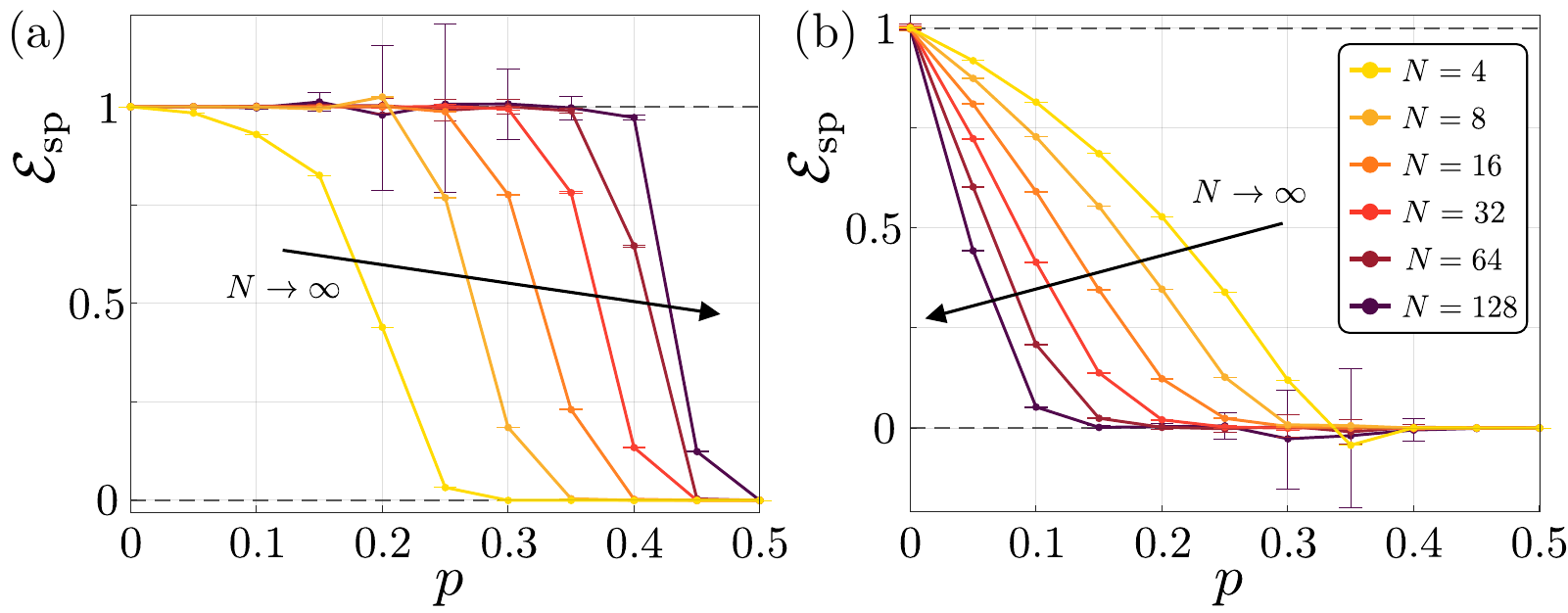}
    \caption{Spurious topological entanglement negativity (TEN), $\EE_{\text{sp}} (N) = \EE_A (2N) - 2\EE_A (N)$, measured in units of $\log 2$, for various system sizes $N$ under (a) $X$-noise and (b) $Z$-noise. Error bars are calculated from $10^6$ to $8\times 10^8$ samples.}
    \label{fig:1DCluster_SpTEN}
\end{figure}

The entanglement negativity $\EE_A = \log \| (\rho_X)^{T_A} \|_1$ for large decohered cluster states can be numerically computed from Eq.~\eqref{1D_TraceNorm} using the Monte Carlo method. We define the spurious TEN as $\EE_{\text{sp}} (N) = \EE_A (2N) - 2\EE_A(N)$, where the arguments in parentheses denote the chain length. Figure~\ref{fig:1DCluster_SpTEN} displays the numerically computed $\EE_{\text{sp}}$ for $\rho_X$ and $\rho_Z$. At $p = 0$ ($p = 1/2$), $\EE_{\text{sp}}$ equals $\log 2$ ($0$) in both cases, consistent with calculations based on the stabilizer formalism~\cite{SM}. For $0 < p < 1/2$, strongly symmetric $\rho_X$ exhibits $\EE_{\text{sp}}$ saturating to $\log 2$ as $N\rightarrow \infty$. In contrast, for weakly symmetric $\rho_Z$, $\EE_{\text{sp}}$ vanishes as $N\rightarrow \infty$.  This suggests that spurious TEN can arise in locally decohered short-range entangled states with strong subsystem symmetry but generally does not appear without such symmetry.

We prove the following theorem to analytically establish the numerically observed connection between strong subsystem symmetry and spurious TEN in 1D:
\begin{theorem*}
Consider a 1D nontrivial mixed-state SPT state $\rho$ protected by on-site $G_1 \times G_2$ symmetry, where $G_i$ are finite Abelian groups acting on separate bipartite subsystems, and let $q$ be the order of the 2-cocycle class associated with the SPT order. Then, $\rho$ has a R\'enyi spurious TEN $\EE_{\text{sp}}^{(2\alpha)} \geq \log q$ for all $\alpha \in \Z_{\geq 2}$ with respect to the bipartition, provided that the matrix product density operator (MPDO) representation of $\rho$ is strongly injective and satisfies condition (C1').
\end{theorem*}
We review the preliminaries and provide the proof of Theorem in SM~\cite{SM}. Here, the R\'enyi-$(2\alpha)$ negativity is defined as $\EE_A^{(2\alpha)} = (2 - 2\alpha)^{-1} \log \left( \Tr[(\rho^{T_A})^{2\alpha}] / \Tr[\rho^{2\alpha}] \right)$, which reduces to $\EE_A$ as $\alpha\rightarrow 1/2$~\cite{calabrese2012entanglement}. The decohered 1D cluster state is a special case of Theorem with $G_{1,2} = \Z_2$ and $q = 2$. The strong injectivity~\cite{xue2024tensor,guo2024locally} and condition (C1') are technical conditions that hold for generic MPDOs representing short-range correlated density matrices. Theorem applies to any 1D nontrivial SPT state decohered under a finite-depth brickwork circuit of symmetry-preserving local Pauli noises with $p < 1/2$~\cite{SM}. (We have $\EE_{\text{sp}}^{(2\alpha)} = \log q$ except for a measure-zero set of mixed states.) Theorem establishes spurious TEN as a universal entanglement quantity that detects mixed anomaly between strong subsystem symmetries in 1D mixed-state $G_1 \times G_2$ SPT orders.

In SM~\cite{SM}, we also show that $X$-decohered 2D cluster states exhibit $\EE_{\text{sp}} = \log 2$ with respect to the bipartition in Fig.~\ref{fig:2DCluster}(c) for $p < 1/2$, capturing the strong subsystem symmetry surviving against SWSSB. This is evident for the $p = 0$ case, where qubits outside the purple region in Fig.~\ref{fig:2DCluster}(c) can be locally disentangled via controlled-$Z$ gates, leaving behind a 1D cluster state along the entangling boundary~\cite{zou2016spurious}. We believe that spurious TEN generally arises for 2D systems with mixed-state SSPT order. Spurious TEN could serve as a signature of a ``mixed-state cluster phase,'' a mixed-state phase two-way connected to the pure cluster state via finite-depth quantum channels, akin to how spurious TEE has been used in the pure-state case~\cite{stephen2019detecting}.

We give two remarks. First, our results demonstrate the existence of spurious TEN in certain locally decohered short-range entangled states. While TEN is commonly used to diagnose mixed-state topological order, these findings highlight that TEN is not invariant under a finite-depth local quantum channel, suggesting caution when using it as a diagnostic tool in general cases. Second, 
Eqs.~\eqref{rhoDX_TA} and \eqref{1D_TraceNorm} also describe the entanglement negativity of the 2D toric code under both $X$ and $Z$ boundary decoherence~\cite{lu2024disentangling}. Our results suggest, as a byproduct, that the long-range entanglement of the 2D toric code remains robust up to the maximal boundary decoherence~\cite{SM}.

%================================%
\textit{Discussion---}In this work, we explore the robustness of 1D and 2D cluster states under local noise preserving strong subsystem symmetry and establish spurious TEN as an information-theoretic quantity that captures the mixed anomaly between strong subsystem symmetries. We discuss how the robustness extends to hold for general cases by employing mappings to stat-mech models and their dimensional reductions. 

While previous studies have discussed the robustness of 1D SPT orders under strongly symmetric open systems~\cite{de2022symmetry,paszko2024edge}, our results contribute further insights by (i) drawing a connection to the modern perspective of SWSSB and (ii) pinpointing the maximal error rate $p = 1/2$ as a mixed-state transition point for both 1D and 2D cluster states under \emph{general} strongly symmetric local Pauli noises. Crucially, our stability arguments do not rely on the replica trick and are free from potential subtleties associated with the R\'enyi fidelity correlators~\cite{lessa2024strong}. Also, we focus on bulk quantities---the fidelity correlator and the spurious TEN---rather than edge modes~\cite{paszko2024edge}.

Our result opens several avenues for future research. First, it has been shown that~\cite{kim2023universal,levin2024physical} for 2D pure states, $\log \mathcal{D}$ provides a universal lower bound for the TEE from the Levin-Wen partition~\cite{levin2006detecting}, where $\mathcal{D}$ is the quantum dimension of the underlying anyon theory. This raises the question of whether a similar lower bound exists for the TEN of mixed states. This could be used to define a \emph{bona fide} diagnostics for mixed-state topological order that remains invariant under finite-depth quantum channels. We note that a convex-roof approach has recently been pursued toward this goal~\cite{wang2025analog}.

Moreover, it would be intriguing to explore connections between mixed-state cluster states and the concept of ``computational phases of matter'' throughout which universal computational power persists in the sense of MBQC. Previous works have linked the computational powers of SPT phases to their group cohomology classification~\cite{stephen2017computational,raussendorf2017symmetry}. Extending this approach to the richer classification of mixed-state SPT orders~\cite{xue2024tensor,guo2024locally} would be an interesting direction. The robustness of mixed-state cluster states found here may hint at the possibility of a ``mixed-state cluster phase'' supporting universal MBQC. Moreover, classifying mixed-state SSPT orders using tensor network techniques would be worthwhile.

%================================%
\textit{Acknowledgements---}We thank Jong Yeon Lee, Cenke Xu, Jintae Kim, and Yun-tak Oh for their helpful discussions. This work was supported by 2021R1A2C4001847, 2022M3H4A1A04074153, National Measurement Standard Services and Technical Services for SME funded by Korea Research Institute of Standards and Science (KRISS – 2024 – GP2024-0015) and the Nano \& Material Technology Development Program through the National Research Foundation of Korea(NRF) funded by Ministry of Science and ICT(RS-2023-00281839).

%================================%
\textit{Data Availability---}The data that support the findings of this article are available from the authors upon reasonable request.

%================================%
\let\oldaddcontentsline\addcontentsline
\renewcommand{\addcontentsline}[3]{}
\bibliographystyle{apsrev4-2}
\bibliography{ref}

@book{wen2004quantum,
  title={Quantum field theory of many-body systems: From the origin of sound to an origin of light and electrons},
  author={Wen, Xiao-Gang},
  year={2004},
  publisher={OUP Oxford}
}

@book{sachdev2023quantum,
  title={Quantum Phases of Matter},
  author={Sachdev, Subir},
  year={2023},
  publisher={Cambridge University Press}
}

@article{levin2006detecting,
  title = {Detecting Topological Order in a Ground State Wave Function},
  author = {Levin, Michael and Wen, Xiao-Gang},
  journal = {Phys. Rev. Lett.},
  volume = {96},
  issue = {11},
  pages = {110405},
  numpages = {4},
  year = {2006},
  month = {Mar},
  publisher = {American Physical Society},
  doi = {10.1103/PhysRevLett.96.110405},
  url = {https://link.aps.org/doi/10.1103/PhysRevLett.96.110405}
}

@article{grover2011entanglement,
  title = {Entanglement entropy of gapped phases and topological order in three dimensions},
  author = {Grover, Tarun and Turner, Ari M. and Vishwanath, Ashvin},
  journal = {Phys. Rev. B},
  volume = {84},
  issue = {19},
  pages = {195120},
  numpages = {13},
  year = {2011},
  month = {Nov},
  publisher = {American Physical Society},
  doi = {10.1103/PhysRevB.84.195120},
  url = {https://link.aps.org/doi/10.1103/PhysRevB.84.195120}
}

@article{cano2015interactions,
  title = {Interactions along an entanglement cut in $2+1\mathrm{D}$ Abelian topological phases},
  author = {Cano, Jennifer and Hughes, Taylor L. and Mulligan, Michael},
  journal = {Phys. Rev. B},
  volume = {92},
  issue = {7},
  pages = {075104},
  numpages = {31},
  year = {2015},
  month = {Aug},
  publisher = {American Physical Society},
  doi = {10.1103/PhysRevB.92.075104},
  url = {https://link.aps.org/doi/10.1103/PhysRevB.92.075104}
}

@article{zou2016spurious,
  title = {Spurious long-range entanglement and replica correlation length},
  author = {Zou, Liujun and Haah, Jeongwan},
  journal = {Phys. Rev. B},
  volume = {94},
  issue = {7},
  pages = {075151},
  numpages = {17},
  year = {2016},
  month = {Aug},
  publisher = {American Physical Society},
  doi = {10.1103/PhysRevB.94.075151},
  url = {https://link.aps.org/doi/10.1103/PhysRevB.94.075151}
}

@article{santos2018symmetry,
  title = {Symmetry-protected topological interfaces and entanglement sequences},
  author = {Santos, Luiz H. and Cano, Jennifer and Mulligan, Michael and Hughes, Taylor L.},
  journal = {Phys. Rev. B},
  volume = {98},
  issue = {7},
  pages = {075131},
  numpages = {22},
  year = {2018},
  month = {Aug},
  publisher = {American Physical Society},
  doi = {10.1103/PhysRevB.98.075131},
  url = {https://link.aps.org/doi/10.1103/PhysRevB.98.075131}
}

@article{williamson2019spurious,
  title = {Spurious Topological Entanglement Entropy from Subsystem Symmetries},
  author = {Williamson, Dominic J. and Dua, Arpit and Cheng, Meng},
  journal = {Phys. Rev. Lett.},
  volume = {122},
  issue = {14},
  pages = {140506},
  numpages = {6},
  year = {2019},
  month = {Apr},
  publisher = {American Physical Society},
  doi = {10.1103/PhysRevLett.122.140506},
  url = {https://link.aps.org/doi/10.1103/PhysRevLett.122.140506}
}

@article{stephen2019detecting,
  title = {Detecting subsystem symmetry protected topological order via entanglement entropy},
  author = {Stephen, David T. and Dreyer, Henrik and Iqbal, Mohsin and Schuch, Norbert},
  journal = {Phys. Rev. B},
  volume = {100},
  issue = {11},
  pages = {115112},
  numpages = {15},
  year = {2019},
  month = {Sep},
  publisher = {American Physical Society},
  doi = {10.1103/PhysRevB.100.115112},
  url = {https://link.aps.org/doi/10.1103/PhysRevB.100.115112}
}

@article{kato2020toy,
  title = {Toy model of boundary states with spurious topological entanglement entropy},
  author = {Kato, Kohtaro and Brand\~ao, Fernando G. S. L.},
  journal = {Phys. Rev. Res.},
  volume = {2},
  issue = {3},
  pages = {032005},
  numpages = {6},
  year = {2020},
  month = {Jul},
  publisher = {American Physical Society},
  doi = {10.1103/PhysRevResearch.2.032005},
  url = {https://link.aps.org/doi/10.1103/PhysRevResearch.2.032005}
}

@article{devakul2018classification,
  title = {Classification of subsystem symmetry-protected topological phases},
  author = {Devakul, Trithep and Williamson, Dominic J. and You, Yizhi},
  journal = {Phys. Rev. B},
  volume = {98},
  issue = {23},
  pages = {235121},
  numpages = {15},
  year = {2018},
  month = {Dec},
  publisher = {American Physical Society},
  doi = {10.1103/PhysRevB.98.235121},
  url = {https://link.aps.org/doi/10.1103/PhysRevB.98.235121}
}

@article{you2018subsystem,
  title = {Subsystem symmetry protected topological order},
  author = {You, Yizhi and Devakul, Trithep and Burnell, F. J. and Sondhi, S. L.},
  journal = {Phys. Rev. B},
  volume = {98},
  issue = {3},
  pages = {035112},
  numpages = {18},
  year = {2018},
  month = {Jul},
  publisher = {American Physical Society},
  doi = {10.1103/PhysRevB.98.035112},
  url = {https://link.aps.org/doi/10.1103/PhysRevB.98.035112}
}

@article{briegel2001persistent,
  title = {Persistent Entanglement in Arrays of Interacting Particles},
  author = {Briegel, Hans J. and Raussendorf, Robert},
  journal = {Phys. Rev. Lett.},
  volume = {86},
  issue = {5},
  pages = {910--913},
  numpages = {0},
  year = {2001},
  month = {Jan},
  publisher = {American Physical Society},
  doi = {10.1103/PhysRevLett.86.910},
  url = {https://link.aps.org/doi/10.1103/PhysRevLett.86.910}
}

@article{raussendorf2001a,
  title = {A One-Way Quantum Computer},
  author = {Raussendorf, Robert and Briegel, Hans J.},
  journal = {Phys. Rev. Lett.},
  volume = {86},
  issue = {22},
  pages = {5188--5191},
  numpages = {0},
  year = {2001},
  month = {May},
  publisher = {American Physical Society},
  doi = {10.1103/PhysRevLett.86.5188},
  url = {https://link.aps.org/doi/10.1103/PhysRevLett.86.5188}
}

@article{raussendorf2003measurement,
  title = {Measurement-based quantum computation on cluster states},
  author = {Raussendorf, Robert and Browne, Daniel E. and Briegel, Hans J.},
  journal = {Phys. Rev. A},
  volume = {68},
  issue = {2},
  pages = {022312},
  numpages = {32},
  year = {2003},
  month = {Aug},
  publisher = {American Physical Society},
  doi = {10.1103/PhysRevA.68.022312},
  url = {https://link.aps.org/doi/10.1103/PhysRevA.68.022312}
}

@article{doherty2009identifying,
  title = {Identifying Phases of Quantum Many-Body Systems That Are Universal for Quantum Computation},
  author = {Doherty, Andrew C. and Bartlett, Stephen D.},
  journal = {Phys. Rev. Lett.},
  volume = {103},
  issue = {2},
  pages = {020506},
  numpages = {4},
  year = {2009},
  month = {Jul},
  publisher = {American Physical Society},
  doi = {10.1103/PhysRevLett.103.020506},
  url = {https://link.aps.org/doi/10.1103/PhysRevLett.103.020506}
}

@article{else2012symmetry1,
  title = {Symmetry-Protected Phases for Measurement-Based Quantum Computation},
  author = {Else, Dominic V. and Schwarz, Ilai and Bartlett, Stephen D. and Doherty, Andrew C.},
  journal = {Phys. Rev. Lett.},
  volume = {108},
  issue = {24},
  pages = {240505},
  numpages = {5},
  year = {2012},
  month = {Jun},
  publisher = {American Physical Society},
  doi = {10.1103/PhysRevLett.108.240505},
  url = {https://link.aps.org/doi/10.1103/PhysRevLett.108.240505}
}

@article{else2012symmetry2,
  title={Symmetry protection of measurement-based quantum computation in ground states},
  author={Else, Dominic V and Bartlett, Stephen D and Doherty, Andrew C},
  journal={New Journal of Physics},
  volume={14},
  number={11},
  pages={113016},
  year={2012},
  publisher={IOP Publishing},
  url={https://iopscience.iop.org/article/10.1088/1367-2630/14/11/113016}
}

@article{stephen2017computational,
  title = {Computational Power of Symmetry-Protected Topological Phases},
  author = {Stephen, David T. and Wang, Dong-Sheng and Prakash, Abhishodh and Wei, Tzu-Chieh and Raussendorf, Robert},
  journal = {Phys. Rev. Lett.},
  volume = {119},
  issue = {1},
  pages = {010504},
  numpages = {5},
  year = {2017},
  month = {Jul},
  publisher = {American Physical Society},
  doi = {10.1103/PhysRevLett.119.010504},
  url = {https://link.aps.org/doi/10.1103/PhysRevLett.119.010504}
}

@article{raussendorf2017symmetry,
  title = {Symmetry-protected topological phases with uniform computational power in one dimension},
  author = {Raussendorf, Robert and Wang, Dong-Sheng and Prakash, Abhishodh and Wei, Tzu-Chieh and Stephen, David T.},
  journal = {Phys. Rev. A},
  volume = {96},
  issue = {1},
  pages = {012302},
  numpages = {14},
  year = {2017},
  month = {Jul},
  publisher = {American Physical Society},
  doi = {10.1103/PhysRevA.96.012302},
  url = {https://link.aps.org/doi/10.1103/PhysRevA.96.012302}
}

@article{raussendorf2019computationally,
  title = {Computationally Universal Phase of Quantum Matter},
  author = {Raussendorf, Robert and Okay, Cihan and Wang, Dong-Sheng and Stephen, David T. and Nautrup, Hendrik Poulsen},
  journal = {Phys. Rev. Lett.},
  volume = {122},
  issue = {9},
  pages = {090501},
  numpages = {5},
  year = {2019},
  month = {Mar},
  publisher = {American Physical Society},
  doi = {10.1103/PhysRevLett.122.090501},
  url = {https://link.aps.org/doi/10.1103/PhysRevLett.122.090501}
}

@article{de2022symmetry,
  title={Symmetry protected topological order in open quantum systems},
  author={de Groot, Caroline and Turzillo, Alex and Schuch, Norbert},
  journal={Quantum},
  volume={6},
  pages={856},
  year={2022},
  publisher={Verein zur F{\"o}rderung des Open Access Publizierens in den Quantenwissenschaften},
  url = {https://quantum-journal.org/papers/q-2022-11-10-856/}
}

@article{lee2022symmetry,
  doi = {10.22331/q-2025-01-23-1607},
  url = {https://doi.org/10.22331/q-2025-01-23-1607},
  title = {Symmetry protected topological phases under decoherence},
  author = {Lee, Jong Yeon and You, Yi-Zhuang and Xu, Cenke},
  journal = {{Quantum}},
  issn = {2521-327X},
  publisher = {{Verein zur F{\"{o}}rderung des Open Access Publizierens in den Quantenwissenschaften}},
  volume = {9},
  pages = {1607},
  month = jan,
  year = {2025}
}

@article{bao2023mixed,
  title={Mixed-state topological order and the errorfield double formulation of decoherence-induced transitions},
  author={Bao, Yimu and Fan, Ruihua and Vishwanath, Ashvin and Altman, Ehud},
  journal={arXiv:2301.05687},
  year={2023},
  url = {https://arxiv.org/abs/2301.05687}
}

@article{lee2023quantum,
  title = {Quantum Criticality Under Decoherence or Weak Measurement},
  author = {Lee, Jong Yeon and Jian, Chao-Ming and Xu, Cenke},
  journal = {PRX Quantum},
  volume = {4},
  issue = {3},
  pages = {030317},
  numpages = {20},
  year = {2023},
  month = {Aug},
  publisher = {American Physical Society},
  doi = {10.1103/PRXQuantum.4.030317},
  url = {https://link.aps.org/doi/10.1103/PRXQuantum.4.030317}
}

@article{fan2024diagnostics,
  title = {Diagnostics of Mixed-State Topological Order and Breakdown of Quantum Memory},
  author = {Fan, Ruihua and Bao, Yimu and Altman, Ehud and Vishwanath, Ashvin},
  journal = {PRX Quantum},
  volume = {5},
  issue = {2},
  pages = {020343},
  numpages = {17},
  year = {2024},
  month = {May},
  publisher = {American Physical Society},
  doi = {10.1103/PRXQuantum.5.020343},
  url = {https://link.aps.org/doi/10.1103/PRXQuantum.5.020343}
}

@article{ma2023average,
  title = {Average Symmetry-Protected Topological Phases},
  author = {Ma, Ruochen and Wang, Chong},
  journal = {Phys. Rev. X},
  volume = {13},
  issue = {3},
  pages = {031016},
  numpages = {24},
  year = {2023},
  month = {Aug},
  publisher = {American Physical Society},
  doi = {10.1103/PhysRevX.13.031016},
  url = {https://link.aps.org/doi/10.1103/PhysRevX.13.031016}
}

@article{sang2024mixed,
  title = {Mixed-State Quantum Phases: Renormalization and Quantum Error Correction},
  author = {Sang, Shengqi and Zou, Yijian and Hsieh, Timothy H.},
  journal = {Phys. Rev. X},
  volume = {14},
  issue = {3},
  pages = {031044},
  numpages = {24},
  year = {2024},
  month = {Sep},
  publisher = {American Physical Society},
  doi = {10.1103/PhysRevX.14.031044},
  url = {https://link.aps.org/doi/10.1103/PhysRevX.14.031044}
}

@article{sang2024stability,
  title={Stability of mixed-state quantum phases via finite Markov length},
  author={Sang, Shengqi and Hsieh, Timothy H},
  journal={arXiv:2404.07251},
  year={2024},
  url = {https://arxiv.org/abs/2310.08639}
}

@article{zou2023channeling,
  title = {Channeling Quantum Criticality},
  author = {Zou, Yijian and Sang, Shengqi and Hsieh, Timothy H.},
  journal = {Phys. Rev. Lett.},
  volume = {130},
  issue = {25},
  pages = {250403},
  numpages = {7},
  year = {2023},
  month = {Jun},
  publisher = {American Physical Society},
  doi = {10.1103/PhysRevLett.130.250403},
  url = {https://link.aps.org/doi/10.1103/PhysRevLett.130.250403}
}

@article{lu2023mixed,
  title = {Mixed-State Long-Range Order and Criticality from Measurement and Feedback},
  author = {Lu, Tsung-Cheng and Zhang, Zhehao and Vijay, Sagar and Hsieh, Timothy H.},
  journal = {PRX Quantum},
  volume = {4},
  issue = {3},
  pages = {030318},
  numpages = {25},
  year = {2023},
  month = {Aug},
  publisher = {American Physical Society},
  doi = {10.1103/PRXQuantum.4.030318},
  url = {https://link.aps.org/doi/10.1103/PRXQuantum.4.030318}
}

@article{guo2024two,
  title = {Two-dimensional symmetry-protected topological phases and transitions in open quantum systems},
  author = {Guo, Yuxuan and Ashida, Yuto},
  journal = {Phys. Rev. B},
  volume = {109},
  issue = {19},
  pages = {195420},
  numpages = {10},
  year = {2024},
  month = {May},
  publisher = {American Physical Society},
  doi = {10.1103/PhysRevB.109.195420},
  url = {https://link.aps.org/doi/10.1103/PhysRevB.109.195420}
}

@article{kuno2024strong,
  title = {Strong-to-weak symmetry breaking states in stochastic dephasing stabilizer circuits},
  author = {Kuno, Yoshihito and Orito, Takahiro and Ichinose, Ikuo},
  journal = {Phys. Rev. B},
  volume = {110},
  issue = {9},
  pages = {094106},
  numpages = {12},
  year = {2024},
  month = {Sep},
  publisher = {American Physical Society},
  doi = {10.1103/PhysRevB.110.094106},
  url = {https://link.aps.org/doi/10.1103/PhysRevB.110.094106}
}

@article{zhang2024strong,
  title={Strong-to-weak spontaneous breaking of 1-form symmetry and intrinsically mixed topological order},
  author={Zhang, Carolyn and Xu, Yichen and Zhang, Jian-Hao and Xu, Cenke and Bi, Zhen and Luo, Zhu-Xi},
  journal={arXiv:2409.17530},
  year={2024},
  url={https://arxiv.org/abs/2409.17530}
}

@article{chen2024strong,
  title={Strong-to-weak Symmetry Breaking and Entanglement Transitions},
  author={Chen, Langxuan and Sun, Ning and Zhang, Pengfei},
  journal={arXiv:2411.05364},
  year={2024},
  url={https://arxiv.org/abs/2411.05364}
}

@article{ma2023topological,
  title={Topological Phases with Average Symmetries: the Decohered, the Disordered, and the Intrinsic},
  author={Ma, Ruochen and Zhang, Jian-Hao and Bi, Zhen and Cheng, Meng and Wang, Chong},
  journal={arXiv:2305.16399},
  year={2023},
  url = {https://arxiv.org/abs/2305.16399}
}

@article{ma2024symmetry,
  title={Symmetry protected topological phases of mixed states in the doubled space},
  author={Ma, Ruochen and Turzillo, Alex},
  journal={arXiv:2403.13280},
  year={2024},
  url = {https://arxiv.org/abs/2403.13280}
}

@article{you2024intrinsic,
  title = {Intrinsic symmetry-protected topological mixed state from modulated symmetries and hierarchical structure of boundary anomaly},
  author = {You, Yizhi and Oshikawa, Masaki},
  journal = {Phys. Rev. B},
  volume = {110},
  issue = {16},
  pages = {165160},
  numpages = {15},
  year = {2024},
  month = {Oct},
  publisher = {American Physical Society},
  doi = {10.1103/PhysRevB.110.165160},
  url = {https://link.aps.org/doi/10.1103/PhysRevB.110.165160}
}

@article{lu2024bilayer,
  title={Bilayer construction for mixed state phenomena with strong, weak symmetries and symmetry breakings},
  author={Lu, S and Zhu, P and Lu, YM},
  journal={arXiv:2411.07174},
  year={2024},
  url={https://arxiv.org/abs/2411.07174}
}

@article{sun2024holographic,
  title={Holographic View of Mixed-State Symmetry-Protected Topological Phases in Open Quantum Systems},
  author={Sun, Shijun and Zhang, Jian-Hao and Bi, Zhen and You, Yizhi},
  journal={arXiv:2410.08205},
  year={2024},
  url={https://arxiv.org/abs/2410.08205}
}

@article{chen2024separability,
  title = {Separability Transitions in Topological States Induced by Local Decoherence},
  author = {Chen, Yu-Hsueh and Grover, Tarun},
  journal = {Phys. Rev. Lett.},
  volume = {132},
  issue = {17},
  pages = {170602},
  numpages = {6},
  year = {2024},
  month = {Apr},
  publisher = {American Physical Society},
  doi = {10.1103/PhysRevLett.132.170602},
  url = {https://link.aps.org/doi/10.1103/PhysRevLett.132.170602}
}

@article{ellison2024towards,
  title={Towards a classification of mixed-state topological orders in two dimensions},
  author={Ellison, Tyler and Cheng, Meng},
  journal={arXiv:2405.02390},
  year={2024},
  url={https://arxiv.org/abs/2405.02390}
}

@article{sohal2024a,
  title={A noisy approach to intrinsically mixed-state topological order},
  author={Sohal, Ramanjit and Prem, Abhinav},
  journal={arXiv:2403.13879},
  year={2024},
  url={https://arxiv.org/abs/2403.13879}
}

@article{chen2023symmetry,
  title = {Symmetry-Enforced Many-Body Separability Transitions},
  author = {Chen, Yu-Hsueh and Grover, Tarun},
  journal = {PRX Quantum},
  volume = {5},
  issue = {3},
  pages = {030310},
  numpages = {32},
  year = {2024},
  month = {Jul},
  publisher = {American Physical Society},
  doi = {10.1103/PRXQuantum.5.030310},
  url = {https://link.aps.org/doi/10.1103/PRXQuantum.5.030310}
}

@article{su2024higher,
  title = {Higher-Form Symmetries under Weak Measurement},
  author = {Su, Kaixiang and Myerson-Jain, Nayan and Wang, Chong and Jian, Chao-Ming and Xu, Cenke},
  journal = {Phys. Rev. Lett.},
  volume = {132},
  issue = {20},
  pages = {200402},
  numpages = {6},
  year = {2024},
  month = {May},
  publisher = {American Physical Society},
  doi = {10.1103/PhysRevLett.132.200402},
  url = {https://link.aps.org/doi/10.1103/PhysRevLett.132.200402}
}

@article{min2024mixed,
  title={Mixed-state phase transitions in spin-Holstein models},
  author={Min, Brett and Zhang, Yuxuan and Guo, Yuxuan and Segal, Dvira and Ashida, Yuto},
  journal={arXiv:2412.02733},
  year={2024},
  url={https://arxiv.org/abs/2412.02733}
}

@article{sala2024stability,
  title={Stability and Loop Models from Decohering Non-Abelian Topological Order},
  author={Sala, Pablo and Verresen, Ruben},
  journal={arXiv:2409.12230},
  year={2024},
  url={https://arxiv.org/abs/2409.12230}
}

@article{shah2024instability,
  title={Instability of steady-state mixed-state symmetry-protected topological order to strong-to-weak spontaneous symmetry breaking},
  author={Shah, Jeet and Fechisin, Christopher and Wang, Yu-Xin and Iosue, Joseph T and Watson, James D and Wang, Yan-Qi and Ware, Brayden and Gorshkov, Alexey V and Lin, Cheng-Ju},
  journal={arXiv:2410.12900},
  year={2024},
  url={https://arxiv.org/abs/2410.12900}
}

@article{zhang2024quantum,
  title={Quantum Communication and Mixed-State Order in Decohered Symmetry-Protected Topological States},
  author={Zhang, Zhehao and Agrawal, Utkarsh and Vijay, Sagar},
  journal={arXiv:2405.05965},
  year={2024},
  url = {https://arxiv.org/abs/2405.05965}
}

@article{lu2024disentangling,
  title = {Disentangling transitions in topological order induced by boundary decoherence},
  author = {Lu, Tsung-Cheng},
  journal = {Phys. Rev. B},
  volume = {110},
  issue = {12},
  pages = {125145},
  numpages = {6},
  year = {2024},
  month = {Sep},
  publisher = {American Physical Society},
  doi = {10.1103/PhysRevB.110.125145},
  url = {https://link.aps.org/doi/10.1103/PhysRevB.110.125145}
}

@article{lee2024mixed,
  title={Mixed-State Topological Order under Coherent Noises},
  author={Lee, Seunghun and Moon, Eun-Gook},
  journal={arXiv:2411.03441},
  year={2024},
  url={https://arxiv.org/abs/2411.03441}
}

@article{li2024replica,
  title={Replica topological order in quantum mixed states and quantum error correction},
  author={Li, Zhuan and Mong, Roger SK},
  journal={arXiv:2402.09516},
  year={2024},
  url = {https://arxiv.org/abs/2401.17359}
}

@article{wang2307intrinsic,
  title={Intrinsic mixed-state quantum topological order},
  author={Wang, Z and Wu, Z and Wang, Z},
  journal={arXiv:2307.13758},
  year={2023},
  url = {https://arxiv.org/abs/2307.13758}
}

@article{lessa2024strong,
  title={Strong-to-weak spontaneous symmetry breaking in mixed quantum states},
  author={Lessa, Leonardo A and Ma, Ruochen and Zhang, Jian-Hao and Bi, Zhen and Cheng, Meng and Wang, Chong},
  journal={arXiv:2405.03639},
  year={2024},
  url = {https://arxiv.org/abs/2405.03639}
}

@article{sala2024spontaneous,
  title = {Spontaneous strong symmetry breaking in open systems: Purification perspective},
  author = {Sala, Pablo and Gopalakrishnan, Sarang and Oshikawa, Masaki and You, Yizhi},
  journal = {Phys. Rev. B},
  volume = {110},
  issue = {15},
  pages = {155150},
  numpages = {28},
  year = {2024},
  month = {Oct},
  publisher = {American Physical Society},
  doi = {10.1103/PhysRevB.110.155150},
  url = {https://link.aps.org/doi/10.1103/PhysRevB.110.155150}
}

@article{gu2024spontaneous,
  title={Spontaneous symmetry breaking in open quantum systems: strong, weak, and strong-to-weak},
  author={Gu, Ding and Wang, Zijian and Wang, Zhong},
  journal={arXiv:2406.19381},
  year={2024},
  url={https://arxiv.org/abs/2406.19381}
}

@article{huang2024hydrodynamics,
  title={Hydrodynamics as the effective field theory of strong-to-weak spontaneous symmetry breaking},
  author={Huang, Xiaoyang and Qi, Marvin and Zhang, Jian-Hao and Lucas, Andrew},
  journal={arXiv:2407.08760},
  year={2024},
  url={https://arxiv.org/abs/2407.08760}
}

@article{liu2024diagnosing,
  title={Diagnosing strong-to-weak symmetry breaking via wightman correlators (2024)},
  author={Liu, Z and Chen, L and Zhang, Y and Zhou, S and Zhang, P},
  journal={arXiv:2410.09327},
  year = {2024},
  url={https://arxiv.org/abs/2410.09327}
}

@article{weinstein2024efficient,
  title={Efficient Detection of Strong-To-Weak Spontaneous Symmetry Breaking via the R{\'e}nyi-1 Correlator (2024)},
  author={Weinstein, Z},
  journal={arXiv:2410.23512},
  year = {2024},
  url={https://arxiv.org/abs/2410.23512}
}

@article{kim2024error,
  title={Error Threshold of SYK Codes from Strong-to-Weak Parity Symmetry Breaking},
  author={Kim, Jaewon and Altman, Ehud and Lee, Jong Yeon},
  journal={arXiv:2410.24225},
  year={2024},
  url={https://arxiv.org/abs/2410.24225}
}

@article{lu2020detecting,
  title = {Detecting Topological Order at Finite Temperature Using Entanglement Negativity},
  author = {Lu, Tsung-Cheng and Hsieh, Timothy H. and Grover, Tarun},
  journal = {Phys. Rev. Lett.},
  volume = {125},
  issue = {11},
  pages = {116801},
  numpages = {6},
  year = {2020},
  month = {Sep},
  publisher = {American Physical Society},
  doi = {10.1103/PhysRevLett.125.116801},
  url = {https://link.aps.org/doi/10.1103/PhysRevLett.125.116801}
}

@article{lu2020structure,
  title = {Structure of quantum entanglement at a finite temperature critical point},
  author = {Lu, Tsung-Cheng and Grover, Tarun},
  journal = {Phys. Rev. Res.},
  volume = {2},
  issue = {4},
  pages = {043345},
  numpages = {13},
  year = {2020},
  month = {Dec},
  publisher = {American Physical Society},
  doi = {10.1103/PhysRevResearch.2.043345},
  url = {https://link.aps.org/doi/10.1103/PhysRevResearch.2.043345}
}

@article{lu2023characterizing,
  title = {Characterizing long-range entanglement in a mixed state through an emergent order on the entangling surface},
  author = {Lu, Tsung-Cheng and Vijay, Sagar},
  journal = {Phys. Rev. Res.},
  volume = {5},
  issue = {3},
  pages = {033031},
  numpages = {16},
  year = {2023},
  month = {Jul},
  publisher = {American Physical Society},
  doi = {10.1103/PhysRevResearch.5.033031},
  url = {https://link.aps.org/doi/10.1103/PhysRevResearch.5.033031}
}

@article{wu2020entanglement,
  title = {Entanglement Renyi Negativity across a Finite Temperature Transition: A Monte Carlo Study},
  author = {Wu, Kai-Hsin and Lu, Tsung-Cheng and Chung, Chia-Min and Kao, Ying-Jer and Grover, Tarun},
  journal = {Phys. Rev. Lett.},
  volume = {125},
  issue = {14},
  pages = {140603},
  numpages = {6},
  year = {2020},
  month = {Sep},
  publisher = {American Physical Society},
  doi = {10.1103/PhysRevLett.125.140603},
  url = {https://link.aps.org/doi/10.1103/PhysRevLett.125.140603}
}

@article{vidal2002computable,
  title = {Computable measure of entanglement},
  author = {Vidal, G. and Werner, R. F.},
  journal = {Phys. Rev. A},
  volume = {65},
  issue = {3},
  pages = {032314},
  numpages = {11},
  year = {2002},
  month = {Feb},
  publisher = {American Physical Society},
  doi = {10.1103/PhysRevA.65.032314},
  url = {https://link.aps.org/doi/10.1103/PhysRevA.65.032314}
}

@article{plenio2005logarithmic,
  title = {Logarithmic Negativity: A Full Entanglement Monotone That is not Convex},
  author = {Plenio, M. B.},
  journal = {Phys. Rev. Lett.},
  volume = {95},
  issue = {9},
  pages = {090503},
  numpages = {4},
  year = {2005},
  month = {Aug},
  publisher = {American Physical Society},
  doi = {10.1103/PhysRevLett.95.090503},
  url = {https://link.aps.org/doi/10.1103/PhysRevLett.95.090503}
}

@article{buvca2012a,
  title={A note on symmetry reductions of the Lindblad equation: transport in constrained open spin chains},
  author={Bu{\v{c}}a, Berislav and Prosen, Toma{\v{z}}},
  journal={New J. Phys.},
  volume={14},
  number={7},
  pages={073007},
  year={2012},
  publisher={IOP Publishing},
  url={https://iopscience.iop.org/article/10.1088/1367-2630/14/7/073007}
}

@article{son2012topological,
  title={Topological order in 1D Cluster state protected by symmetry},
  author={Son, Wonmin and Amico, Luigi and Vedral, Vlatko},
  journal={Quantum Inf. Process.},
  volume={11},
  pages={1961--1968},
  year={2012},
  publisher={Springer},
  url={https://link.springer.com/article/10.1007/s11128-011-0346-7}
}

@article{chen2014symmetry,
  title={Symmetry-protected topological phases from decorated domain walls},
  author={Chen, Xie and Lu, Yuan-Ming and Vishwanath, Ashvin},
  journal={Nat. Commun.},
  volume={5},
  number={1},
  pages={3507},
  year={2014},
  publisher={Nature Publishing Group UK London},
  url={https://www.nature.com/articles/ncomms4507}
}

@article{kim2023universal,
  title = {Universal Lower Bound on Topological Entanglement Entropy},
  author = {Kim, Isaac H. and Levin, Michael and Lin, Ting-Chun and Ranard, Daniel and Shi, Bowen},
  journal = {Phys. Rev. Lett.},
  volume = {131},
  issue = {16},
  pages = {166601},
  numpages = {6},
  year = {2023},
  month = {Oct},
  publisher = {American Physical Society},
  doi = {10.1103/PhysRevLett.131.166601},
  url = {https://link.aps.org/doi/10.1103/PhysRevLett.131.166601}
}

@article{levin2024physical,
  title = {Physical proof of the topological entanglement entropy inequality},
  author = {Levin, Michael},
  journal = {Phys. Rev. B},
  volume = {110},
  issue = {16},
  pages = {165154},
  numpages = {11},
  year = {2024},
  month = {Oct},
  publisher = {American Physical Society},
  doi = {10.1103/PhysRevB.110.165154},
  url = {https://link.aps.org/doi/10.1103/PhysRevB.110.165154}
}

@article{kim2024persistent,
  title={Persistent Topological Negativity in a High-Temperature Mixed-State},
  author={Kim, Yonna and Lavasani, Ali and Vijay, Sagar},
  journal={arXiv:2408.00066},
  year={2024},
  url = {https://arxiv.org/abs/2408.00066}
}

@article{verresen2021efficiently,
  title={Efficiently preparing Schr$\backslash$" odinger's cat, fractons and non-Abelian topological order in quantum devices},
  author={Verresen, Ruben and Tantivasadakarn, Nathanan and Vishwanath, Ashvin},
  journal={arXiv:2112.03061},
  year={2021},
  url={https://arxiv.org/abs/2112.03061}
}

@article{lu2022measurement,
  title = {Measurement as a Shortcut to Long-Range Entangled Quantum Matter},
  author = {Lu, Tsung-Cheng and Lessa, Leonardo A. and Kim, Isaac H. and Hsieh, Timothy H.},
  journal = {PRX Quantum},
  volume = {3},
  issue = {4},
  pages = {040337},
  numpages = {22},
  year = {2022},
  month = {Dec},
  publisher = {American Physical Society},
  doi = {10.1103/PRXQuantum.3.040337},
  url = {https://link.aps.org/doi/10.1103/PRXQuantum.3.040337}
}

@article{lee2022decoding,
  title={Decoding Measurement-Prepared Quantum Phases and Transitions: from Ising model to gauge theory, and beyond},
  author={Lee, Jong Yeon and Ji, Wenjie and Bi, Zhen and Fisher, Matthew},
  journal={arXiv:2208.11699},
  year={2022},
  url={https://arxiv.org/abs/2208.11699}
}

@article{zhu2023nishimoris,
  title = {Nishimori's Cat: Stable Long-Range Entanglement from Finite-Depth Unitaries and Weak Measurements},
  author = {Zhu, Guo-Yi and Tantivasadakarn, Nathanan and Vishwanath, Ashvin and Trebst, Simon and Verresen, Ruben},
  journal = {Phys. Rev. Lett.},
  volume = {131},
  issue = {20},
  pages = {200201},
  numpages = {9},
  year = {2023},
  month = {Nov},
  publisher = {American Physical Society},
  doi = {10.1103/PhysRevLett.131.200201},
  url = {https://link.aps.org/doi/10.1103/PhysRevLett.131.200201}
}

@article{tantivasadakarn2023hierarchy,
  title = {Hierarchy of Topological Order From Finite-Depth Unitaries, Measurement, and Feedforward},
  author = {Tantivasadakarn, Nathanan and Vishwanath, Ashvin and Verresen, Ruben},
  journal = {PRX Quantum},
  volume = {4},
  issue = {2},
  pages = {020339},
  numpages = {25},
  year = {2023},
  month = {Jun},
  publisher = {American Physical Society},
  doi = {10.1103/PRXQuantum.4.020339},
  url = {https://link.aps.org/doi/10.1103/PRXQuantum.4.020339}
}

@article{tantivasadakarn2023shortest,
  title = {Shortest Route to Non-Abelian Topological Order on a Quantum Processor},
  author = {Tantivasadakarn, Nathanan and Verresen, Ruben and Vishwanath, Ashvin},
  journal = {Phys. Rev. Lett.},
  volume = {131},
  issue = {6},
  pages = {060405},
  numpages = {5},
  year = {2023},
  month = {Aug},
  publisher = {American Physical Society},
  doi = {10.1103/PhysRevLett.131.060405},
  url = {https://link.aps.org/doi/10.1103/PhysRevLett.131.060405}
}

@article{ding2024boundary,
  title={Boundary anomaly detection in two-dimensional subsystem symmetry-protected topological phases},
  author={Ding, Ke and Zhang, Hao-Ran and Liu, Bai-Ting and Yang, Shuo},
  journal={arXiv preprint arXiv:2412.07563},
  year={2024},
  url = {https://arxiv.org/abs/2412.07563}
}

@misc{one,
  author = {}, 
  title = {}, 
  year = {}, 
  note = {In terms of the doubled Hilbert-space formalism, SWSSB corresponds to spontaneous breaking of $G \times G$ symmetry of the doubled state (strong symmetry) down to its diagonal $G$ symmetry (weak symmetry)~\cite{lee2023quantum}.}
}

@misc{two,
  author = {}, 
  title = {}, 
  year = {}, 
  note = {$x \in S \triangle S'$ if and only if $x \in S\setminus S'$ or $x \in S'\setminus S$.}
}

@misc{SM,
  author = {}, 
  title = {}, 
  year = {}, 
  note = {See the Supplementary Material (SM), which includes Refs.~\cite{lu2020detecting,lu2023characterizing,lu2024disentangling,zou2016spurious,shi2020entanglement,sang2021entanglement,calabrese2012entanglement,mueller2017exact,schollwock2011density,perezgarcia2007matrix,perezgarcia2008string,chen2011classification,ma2023average,xue2024tensor,guo2024locally,verstraete2004matrix,duivenvoorden2017entanglement,lee2025toappear}, for detailed derivations of fidelity correlators and entanglement negativity for 1D and 2D cluster states, the proof of Theorem, and a connection with the boundary decoherence problem in 2D toric code.}
}

@misc{three,
  author = {}, 
  title = {}, 
  year = {}, 
  note = {For a 1D cluster state decohered under $X$-noise, it suffices to sum over even-sized subsets $S$ of the sublattice $A$, as the contributions from sublattice $B$ cancel between the numerator and denominator of Eq.~\eqref{1D_FC}.}
}

@misc{four,
  author = {}, 
  title = {}, 
  year = {}, 
  Note = {The same asymptotic behavior of the fidelity correlator has been independently derived in Refs. \cite{weinstein2024efficient} for a different setting involving the $\Z_2$-symmetric product state under a $ZZ$-dephasing channel.}
}

@misc{five,
  author = {}, 
  title = {}, 
  year = {}, 
  Note = {On a finite-height cylinder, the $ZXZ$ stabilizers at the top and bottom boundaries introduce additional boundary interactions for $\sigma$ spins in Eq.~\eqref{2D_rhoX}, whose effect becomes negligible for $p < 1/2$ in the limit of large height. See SM~\cite{SM} for a detailed discussion.} 
}

@article{calabrese2012entanglement,
  title = {Entanglement Negativity in Quantum Field Theory},
  author = {Calabrese, Pasquale and Cardy, John and Tonni, Erik},
  journal = {Phys. Rev. Lett.},
  volume = {109},
  issue = {13},
  pages = {130502},
  numpages = {5},
  year = {2012},
  month = {Sep},
  publisher = {American Physical Society},
  doi = {10.1103/PhysRevLett.109.130502},
  url = {https://link.aps.org/doi/10.1103/PhysRevLett.109.130502}
}

@article{mueller2017exact,
title = {Exact solutions to plaquette Ising models with free and periodic boundaries},
journal = {Nucl. Phys. B},
volume = {914},
pages = {388-404},
year = {2017},
issn = {0550-3213},
doi = {https://doi.org/10.1016/j.nuclphysb.2016.11.005},
url = {https://www.sciencedirect.com/science/article/pii/S0550321316303558},
author = {Marco Mueller and Desmond A. Johnston and Wolfhard Janke}
}

@article{xue2024tensor,
  title={Tensor network formulation of symmetry protected topological phases in mixed states},
  author={Xue, Hanyu and Lee, Jong Yeon and Bao, Yimu},
  journal={arXiv:2403.17069},
  year={2024},
  url={https://arxiv.org/abs/2403.17069}
}

@article{guo2024locally,
  title={Locally purified density operators for symmetry-protected topological phases in mixed states},
  author={Guo, Yuchen and Zhang, Jian-Hao and Zhang, Hao-Ran and Yang, Shuo and Bi, Zhen},
  journal={arXiv:2403.16978},
  year={2024},
  url={https://arxiv.org/abs/2403.16978}
}

@article{bathas1995two,
  title={Two-and three-dimensional spin systems with gonihedric action},
  author={Bathas, GK and Floratos, E and Savvidy, George K and Savvidy, Konstantin G},
  journal={Mod. Phys. Lett. A},
  volume={10},
  number={35},
  pages={2695--2701},
  year={1995},
  publisher={World Scientific},
  url={https://www.worldscientific.com/doi/abs/10.1142/S0217732395002829}
}

@article{paszko2024edge,
  title = {Edge Modes and Symmetry-Protected Topological States in Open Quantum Systems},
  author = {Paszko, Dawid and Rose, Dominic C. and Szyma\ifmmode \acute{n}\else \'{n}\fi{}ska, Marzena H. and Pal, Arijeet},
  journal = {PRX Quantum},
  volume = {5},
  issue = {3},
  pages = {030304},
  numpages = {22},
  year = {2024},
  month = {Jul},
  publisher = {American Physical Society},
  doi = {10.1103/PRXQuantum.5.030304},
  url = {https://link.aps.org/doi/10.1103/PRXQuantum.5.030304}
}

@article{shi2020entanglement,
  title={Entanglement negativity at the critical point of measurement-driven transition},
  author={Shi, Bowen and Dai, Xin and Lu, Yuan-Ming},
  journal={arXiv:2012.00040},
  year={2020},
  url={https://arxiv.org/abs/2012.00040}
}

@article{sang2021entanglement,
  title = {Entanglement Negativity at Measurement-Induced Criticality},
  author = {Sang, Shengqi and Li, Yaodong and Zhou, Tianci and Chen, Xiao and Hsieh, Timothy H. and Fisher, Matthew P.A.},
  journal = {PRX Quantum},
  volume = {2},
  issue = {3},
  pages = {030313},
  numpages = {23},
  year = {2021},
  month = {Jul},
  publisher = {American Physical Society},
  doi = {10.1103/PRXQuantum.2.030313},
  url = {https://link.aps.org/doi/10.1103/PRXQuantum.2.030313}
}

@article{schollwock2011density,
  title={The density-matrix renormalization group in the age of matrix product states},
  author={Schollw{\"o}ck, Ulrich},
  journal={Ann. Phys.},
  volume={326},
  number={1},
  pages={96--192},
  year={2011},
  publisher={Elsevier},
  url={https://www.sciencedirect.com/science/article/abs/pii/S0003491610001752}
}

@article{perezgarcia2007matrix,
  title={Matrix product state representations},
  author={Perez-Garcia, David and Verstraete, Frank and Wolf, Michael M and Cirac, J Ignacio},
  journal={Quantum Inf. Comput.},
  volume = {7},
  pages = {401},
  year={2007},
  url={https://dl.acm.org/doi/10.5555/2011832.2011833}
}

@article{perezgarcia2008string,
  title = {String Order and Symmetries in Quantum Spin Lattices},
  author = {P\'erez-Garc\'{\i}a, D. and Wolf, M. M. and Sanz, M. and Verstraete, F. and Cirac, J. I.},
  journal = {Phys. Rev. Lett.},
  volume = {100},
  issue = {16},
  pages = {167202},
  numpages = {4},
  year = {2008},
  month = {Apr},
  publisher = {American Physical Society},
  doi = {10.1103/PhysRevLett.100.167202},
  url = {https://link.aps.org/doi/10.1103/PhysRevLett.100.167202}
}

@article{chen2011classification,
  title = {Classification of gapped symmetric phases in one-dimensional spin systems},
  author = {Chen, Xie and Gu, Zheng-Cheng and Wen, Xiao-Gang},
  journal = {Phys. Rev. B},
  volume = {83},
  issue = {3},
  pages = {035107},
  numpages = {19},
  year = {2011},
  month = {Jan},
  publisher = {American Physical Society},
  doi = {10.1103/PhysRevB.83.035107},
  url = {https://link.aps.org/doi/10.1103/PhysRevB.83.035107}
}

@article{verstraete2004matrix,
  title = {Matrix Product Density Operators: Simulation of Finite-Temperature and Dissipative Systems},
  author = {Verstraete, F. and Garc\'{\i}a-Ripoll, J. J. and Cirac, J. I.},
  journal = {Phys. Rev. Lett.},
  volume = {93},
  issue = {20},
  pages = {207204},
  numpages = {4},
  year = {2004},
  month = {Nov},
  publisher = {American Physical Society},
  doi = {10.1103/PhysRevLett.93.207204},
  url = {https://link.aps.org/doi/10.1103/PhysRevLett.93.207204}
}

@article{duivenvoorden2017entanglement,
  title = {Entanglement phases as holographic duals of anyon condensates},
  author = {Duivenvoorden, Kasper and Iqbal, Mohsin and Haegeman, Jutho and Verstraete, Frank and Schuch, Norbert},
  journal = {Phys. Rev. B},
  volume = {95},
  issue = {23},
  pages = {235119},
  numpages = {24},
  year = {2017},
  month = {Jun},
  publisher = {American Physical Society},
  doi = {10.1103/PhysRevB.95.235119},
  url = {https://link.aps.org/doi/10.1103/PhysRevB.95.235119}
}

@article{wang2025analog,
  title = {Analog of Topological Entanglement Entropy for Mixed States},
  author = {Wang, Ting-Tung and Song, Menghan and Meng, Zi Yang and Grover, Tarun},
  journal = {PRX Quantum},
  volume = {6},
  issue = {1},
  pages = {010358},
  numpages = {23},
  year = {2025},
  month = {Mar},
  publisher = {American Physical Society},
  doi = {10.1103/PRXQuantum.6.010358},
  url = {https://link.aps.org/doi/10.1103/PRXQuantum.6.010358}
}

@article{guo2025quantum,
  title={Quantum Strong-to-Weak Spontaneous Symmetry Breaking in Decohered Critical Spin Chain},
  author={Guo, Yuxuan and Yang, Sheng and Yu, Xue-Jia},
  journal={arXiv preprint arXiv:2503.14221},
  year={2025},
  url = {https://arxiv.org/abs/2503.14221}
}

@article{coser2019classification,
  doi = {10.22331/q-2019-08-12-174},
  url = {https://doi.org/10.22331/q-2019-08-12-174},
  title = {Classification of phases for mixed states via fast dissipative evolution},
  author = {Coser, Andrea and P{\'{e}}rez-Garc{\'{i}}a, David},
  journal = {{Quantum}},
  issn = {2521-327X},
  publisher = {{Verein zur F{\"{o}}rderung des Open Access Publizierens in den Quantenwissenschaften}},
  volume = {3},
  pages = {174},
  month = aug,
  year = {2019}
}

@misc{lee2025toappear,
  author = {}, 
  title = {}, 
  year = {}, 
  note = {S. Lee, M. Kim \& E.-G. Moon, (to appear).}
}
\let\addcontentsline\oldaddcontentsline

%================================%
\onecolumngrid

\renewcommand{\thefigure}{S\arabic{figure}}
\renewcommand{\theequation}{S\arabic{equation}}
\renewcommand{\thetable}{S\Roman{table}}
\renewcommand{\thesection}{S\Roman{section}}
\renewcommand{\thesubsection}{\Alph{subsection}}
\renewcommand{\thesubsubsection}{\arabic{subsubsection}}

\newpage

\centerline{\large{\textbf{Supplementary Material: Robust Mixed-State Cluster States and}}}
\vspace{3pt}
\centerline{\large{\textbf{Spurious Topological Entanglement Negativity}}}
\vspace{10pt}

\centerline{Seunghun Lee and Eun-Gook Moon}
\centerline{\textit{Department of Physics, Korea Advanced Institute of Science and Technology, Daejeon 34141, Republic of Korea}}

\setcounter{equation}{0}
\setcounter{figure}{0}
\setcounter{secnumdepth}{3}
\setcounter{theorem}{0}
\setcounter{lemma}{0}
\tableofcontents

\section{1D Noisy Cluster States}

In this Section, we consider the decohered 1D cluster state $\rho_0 = | \psi_0 \rangle \langle \psi_0 |$ on a periodic chain of $2N$ qubits. In Sec.~\ref{Sec:1D_FC}, the fidelity correlator of the 1D decohered cluster states and the stability of their mixed-state SPT order are discussed. In Sec.~\ref{Sec:1D_EN}, the entanglement negativity of decohered cluster states and its stat-mech expression is explained. In Sec.~\ref{Sec:1D_Thm1}, a proof of Theorem in the main text is provided.

\subsection{Fidelity Correlator} \label{Sec:1D_FC}

In Sec.~\ref{Sec:1D_FC:X}, we exactly compute the fidelity correlator of the $X$-decohered 1D cluster state and demonstrate its exponential decay for $p < 1/2$. In Sec.~\ref{Sec:1D_FC:General}, we compute the fidelity correlator under general local Pauli noise and show the stability of the mixed-state SPT order of the decohered cluster state. In Sec.~\ref{Sec:1D_FC:NonPauli}, we provide numerical evidence of the robustness of the mixed-state SPT order under strongly symmetric non-Pauli noise. 

\subsubsection{Fidelity Correlator for \texorpdfstring{$X$}{X}-Noise} \label{Sec:1D_FC:X}

The stat-mech expression for the fidelity correlator of the strongly symmetric $\rho_X$ is given by Eq.~\eqref{1D_FC} in the main text, which we repeat here:
\begin{align} \label{1D_FC_SM}
    F_Z (x,y) = \frac{\sum_S \left[ \langle \sigma_S \rangle_\beta \langle \sigma_{S \triangle \{x,y\}} \rangle_\beta \right]^{1/2}}{\sum_S \langle \sigma_S \rangle_\beta},
\end{align}
where the summation runs over all even-sized subsets $S$ of the odd sublattice $A$ (see the footnote~\cite{two}), $\sigma_S \equiv \prod_{\in S} \sigma_i$, $\triangle$ represents the symmetric difference, $\langle \cdot \rangle_\beta = Z_\beta^{-1} \sum_{\{\sigma\}} (\cdot) e^{\beta \sum_{i=1}^N \sigma_i \sigma_{i+1}}$ denotes the expectation value in the 1D Ising model at inverse temperature $\beta = -\frac 12 \log (1 - 2p)$, and $Z_\beta = 2^N [(\sinh\beta)^N + (\cosh\beta)^N]$ is the corresponding partition function. Here, we exactly compute the fidelity correlator in Eq.~\eqref{1D_FC_SM} for finite $N$ and examine its asymptotic behavior in the thermodynamic limit.

First, the denominator of Eq.~\eqref{1D_FC_SM} can be easily computed as follows:
\begin{align}
    \sum_{S} \langle \sigma_S \rangle_\beta = \left\langle \prod_{j=1}^N (1 + \sigma_j) \right\rangle_\beta = 2^N \left\langle \prod_{j=1}^N \delta_{\sigma_j, 1} \right\rangle_\beta = \frac{e^{\beta N}}{(\sinh\beta)^N + (\cosh\beta)^N}.
\end{align}

For the numerator, let's simplify an expression for the multi-spin correlator $\langle \sigma_S \rangle_\beta$. Using the transfer matrix of the 1D Ising model given by $T = \begin{pmatrix} e^\beta & e^{-\beta} \\ e^{-\beta} & e^\beta \end{pmatrix} = e^\beta I + e^{-\beta} X$, it can be rewritten as
\begin{align}
    \langle \sigma_S \rangle_\beta = \frac{1}{Z_\beta} \Tr \left[ \prod_{j=1}^N (e^\beta I + e^{-\beta} X) Z^{s_j} \right],
\end{align}
where $\mathbf{s} = \{s_{2j-1}\}_{j=1}^N \in \{0,1\}^N$ represents the support of subsets $S$ of the odd sublattice $A$ and satisfies the constraint $\sum_{j=1}^N s_{2j-1} \in 2\Z$. Any such $\mathbf{s}$ can be one-to-two mapped to a new bit-string $\mathbf{e} = \{e_{2j}\}_{j=1}^N \in \{0,1\}^N$, with each $e_{2j}$ representing a site in the even sublattice $B$ between sites $2j-1$ and $2j+1$ in the sublattice $A$. (In the context of quantum error correction, $\mathbf{s}$ represents the error syndrome and $\mathbf{e}$ the location of Pauli-$X$ Kraus operators on the sublattice $B$.) Since $Z(e^{\beta} I + e^{-\beta} X)^l Z = (e^{\beta} I - e^{-\beta} X)^l$, this observation leads to $\langle \sigma_S \rangle_\beta = Z_\beta^{-1} \Tr \left[ (e^{\beta} I + e^{-\beta} X)^{N - |\mathbf{e}|} (e^{\beta} I - e^{-\beta} X)^{|\mathbf{e}|} \right]$, where $|\mathbf{e}| = \sum_{j=1}^N e_{2j}$. Now, from the relations $e^\beta + e^{-\beta} X = V \Lambda V^\dagger$ and $e^\beta - e^{-\beta} X = V^\dagger \Lambda V$ with
\begin{align}
    \Lambda = \begin{pmatrix}
        2\sinh\beta & 0 \\
        0 & 2\cosh\beta 
    \end{pmatrix}, \qquad V = \frac{1}{\sqrt{2}} \begin{pmatrix}
        1 & 1 \\
        -1 & 1
    \end{pmatrix},
\end{align}
we obtain
\begin{align}
    \langle \sigma_S \rangle_\beta = \frac{1}{Z_\beta} \Tr \left[ \Lambda^{N - |\mathbf{e}|} (V^\dagger)^2 \Lambda^{|\mathbf{e}|} V^2 \right] = \frac{(\tanh\beta)^{N - |\mathbf{e}|} + (\tanh\beta)^{|\mathbf{e}|}}{1 + (\tanh\beta)^N}.
\end{align}

For $\langle \sigma_{S \triangle \{x,y\}} \rangle_\beta$, the corresponding bit-string is given by $\mathbf{e} \oplus \mathbf{r}^{(x,y)}$, where $\oplus$ is a vector addition modulo 2 and $\mathbf{r}^{(x,y)} = \{r_{2j}^{(x,y)}\}_{j=1}^N  \in \{0,1\}^N$ is defined as $r_{2j}^{(x,y)} = 1$ for $2j \in (x, y)$ and $0$ otherwise. Consequently, 
\begin{align}
    F_Z(x,y) = \frac{\sum_{\mathbf{e}} \sqrt{(\tanh\beta)^{N - |\mathbf{e}|} + (\tanh\beta)^{|\mathbf{e}|}} \sqrt{ (\tanh\beta)^{N - |\mathbf{e} \oplus \mathbf{r}^{(x,y)}|} + (\tanh\beta)^{|\mathbf{e} \oplus \mathbf{r}^{(x,y)}|} }}{2 e^{\beta N} (\mathrm{sech} \beta)^N},
\end{align}
where the factor $2$ in the denominator is from the one-to-two nature of the mapping from $\mathbf{s}$ to $\mathbf{e}$. Letting $n_1$ ($m_1$) be the number of indices $2j \in (x, y)$ ($2j \not\in (x, y)$) with $e_{2j} = 1$, we have $|\mathbf{e}| = n_1 + m_1$ and $|\mathbf{e} \oplus \mathbf{r}^{(x,y)}| = \frac 12 |x-y| - n_1 + m_1$. Therefore, we obtain the following exact expression for finite $N$: 
\begin{equation}
    \begin{aligned} \label{1D_FC_Exact}
        F_Z(|x-y|) &= \frac{(\cosh\beta)^N}{2 e^{\beta N}} \sum_{n_1 = 0}^{N/2} \binom{|x-y|/2}{n_1} \sum_{m_1 = 0}^{N/2} \binom{N - |x-y|/2}{m_1} \,f(|x-y|, n_1, m_1), \\
        f(|x-y|, n_1, m_1) &= \sqrt{ (\tanh\beta)^{N-n_1-m_1} + (\tanh\beta)^{n_1+m_1} } \sqrt{ (\tanh\beta)^{N-\frac 12 |x-y|+n_1-m_1} + (\tanh\beta)^{\frac 12 |x-y|-n_1+m_1} }.
    \end{aligned}
\end{equation}

Now, let $N$ be even for simplicity, and take $x = 1$ and $y = N + 1$, which correspond to the largest separation between charged operators $Z_{x,y}$. For large $N$, we can approximate a binomial coefficient by a Gaussian distribution as $2^{-N/2} \binom{N/2}{n_1} \simeq (\pi N/4)^{-1/2} \exp \left[ -\frac{(n_1 - N/4)^2}{N/4} \right]$, and similarly for $\binom{N/2}{m_1}$. Since $f(|x-y|, n_1, m_1)$ is upper bounded by $2$ and the Gaussian distributions are sharply peaked around $N/4$ in the thermodynamic limit, the asymptotic behavior of $F_Z(x,y)$ can be obtained by replacing the summations in Eq.~\eqref{1D_FC_Exact} by $f(N, N/4, N/4) = 2 (\tanh\beta)^{N/2}$: 
\begin{align} \label{FC_exp}
    F_Z(|x-y| = N) \,\underset{N\rightarrow \infty}{\longrightarrow}\, \frac{(\cosh\beta)^N \cdot 2^N \cdot 2 (\tanh\beta)^{N/2}}{2 e^{\beta N}} = e^{-N / \xi},
\end{align}
where $\xi = -1 / \log [2\sqrt{p(1-p)}]$. Therefore, the fidelity correlator of $\rho_X$ decays exponentially for $p < 1/2$, where the decaying length scale $\xi$ is finite. On the other hand, at $p = 1/2$, it is easy to see that $F_Z(|x-y|) = 1$ for any $N$. 

\subsubsection{Fidelity Correlator for General Pauli Noises} \label{Sec:1D_FC:General}

For a general local incoherent Pauli noise $\NN_j^P [\rho] = (1-p) \rho + p P_j \rho P_j$ ($P_j$ can act on a finite number of qubits near site $j$) that respects strong subsystem symmetry, we can similarly argue the stability of the mixed-state SPT order of the decohered 1D cluster state $\rho_P = \prod_{j=1}^{2N} \NN_j^P [\rho_0]$ up to the maximal error rate $p = 1/2$. 

First, the stat-mech model arising in the spectrum of the decohered density matrix is local and subsystem symmetric. This follows from $[P_j, G_{A/B}] = 0$, where locality ensures $P_j$ is a product of finitely many $X$-operators and an even number of $Z$-operators in both $A$ and $B$ sublattices. As discussed in the main text, conjugation by $X_i$ introduces a two-spin interaction between neighboring Ising spins $\sigma_{i \pm 1}$, while conjugation by $Z_i$ multiplies $\sigma_i$ to the existing interaction. Thus, the resulting stat-mech model for the fidelity correlator contains local interactions $\sigma_{U_j} = \prod_{i\in U_j} \sigma_i$ that respect the subsystem symmetries, i.e., $U_i$ has an even number of sites per sublattice. 
The spectrum of the decohered density matrix can then be expressed as
\begin{align} \label{spec_P}
    \rho_P (\{K\}) \propto \sum_{\{a\}} \prod_{j=1}^{2N} K_j^{a_j} e^{\beta \sum_{i=1}^{2N} \sigma_{U_i}},
\end{align}
where $\beta = -\frac 12 \log (1-2p)$.

Next, consider the spectrum of $O_x O_y^\dagger \rho_P O_x^\dagger O_y$ for charged local Pauli operators $O_{x,y}$. Since $K_j |\psi_0 \rangle = | \psi_0 \rangle$, conjugating $\rho_P$ by $O_x$ and $O_y^\dagger$ is equivalent to conjugating it by $\prod_{i\in R_x \cup R_y} Z_i$, where $R_{x,y}$ are some regions near sites $x$ and $y$, respectively. Combining this with Eq.~\eqref{spec_P}, we obtain
\begin{align} \label{FC_O}
    F_O (x,y) = \frac{\sum_S \left[ \langle\!\langle \sigma_S \rangle\!\rangle_\beta \langle\!\langle \sigma_{S \triangle (R_x \cup R_y)} \rangle\!\rangle_\beta \right]^{1/2}}{\sum_S \langle\!\langle \sigma_S \rangle\!\rangle_\beta},
\end{align}
where $\langle\!\langle \cdot \rangle\!\rangle_\beta$ represents the expectation value for the stat-mech model $H = -\sum_{i=1}^{2N} \sigma_{U_i}$ at inverse temperature $\beta$. Here, $S$ runs over all subsets that are modulo-2 union of local patches $U_i$ i.e., $S = \bigoplus_{j=1}^{2N} U_i^{e_i}$ with $e_i \in \{0,1\}$. (These subsets $S$ correspond to error syndromes resulting from the stochastic action of Kraus operators $P_i$ on $\rho_0$.) To ensure that Eq.~\eqref{FC_O} does not vanish trivially for all $p$, the operators $O_{x,y}$ must satisfy $R_x \cup R_y = \bigoplus_{i=1}^{2N} U_i^{r_i}$ for some $r_i \in \{0,1\}$. (The number of $i$ with $r_i = 1$ increases as $|x - y| \rightarrow \infty$.) Henceforth, we assume such choices of $O_{x,y}$.

Since $U_i$ has an even number of sites per sublattice, there are an even number of $U_i$ containing a certain site $j$ in sublattice $A$, and similarly for sublattice $B$. Thus, $\prod_{i=1}^N \sigma_{U_{2i-1}} = \prod_{i=1}^N \sigma_{U_{2i}} = 1$. Now, let's introduce a new spin variables $\tau_i$ as $\sigma_{U_i} = \tau_{i-1} \tau_{i+1}$ ($\tau_i \in \{\pm 1\}$), which is valid since $\prod_{i=1}^N \sigma_{U_{2i-1}} = \prod_{i=1}^N \tau_{2i-2} \tau_{2i} = 1$ and $\prod_{i=1}^N \sigma_{U_{2i}} = \prod_{i=1}^N \tau_{2i-1} \tau_{2i+1} = 1$. Using this, we have
\begin{align}
    \langle\!\langle \sigma_S \rangle\!\rangle_\beta = \frac{\sum_{\{\sigma\}} \sigma_S e^{\beta \sum_{i=1}^{2N} \sigma_{U_i}}}{\sum_{\{\sigma\}} e^{\beta \sum_{i=1}^{2N} \sigma_{U_i}}} = \frac{\sum_{\{\tau\}} \prod_{j=1}^{2N} (\tau_j \tau_{j+2})^{e_j} e^{\beta \sum_{i=1}^{2N} \tau_i \tau_{i+2}}}{\sum_{\{\tau\}} e^{\beta \sum_{i=1}^{2N} \tau_i \tau_{i+2}}} = \langle \tau_{S_A} \rangle_\beta \langle \tau_{S_B} \rangle_\beta,
\end{align}
where $S_A$ ($S_B$) is the subset of sublattice $A$ ($B$) that corresponds to the bit-string $\{e_{2j}\}_{j=1}^N$ ($\{e_{2j-1}\}_{j=1}^N$), and $\langle \cdot \rangle_\beta$ is the expectation value in the 1D Ising model of length $N$. Similarly,
\begin{align}
    \langle\!\langle \sigma_{S \triangle (R_x \cup R_y)} \rangle\!\rangle_\beta = \frac{\sum_{\{\tau\}} \prod_{j=1}^{2N} (\tau_j \tau_{j+2})^{e_j \oplus r_j} e^{\beta \sum_{i=1}^{2N} \tau_i \tau_{i+2}}}{\sum_{\{\tau\}} e^{\beta \sum_{i=1}^{2N} \tau_i \tau_{i+2}}} = \langle \tau_{S_A \triangle R_A} \rangle_\beta \langle \tau_{S_B \triangle R_B} \rangle_\beta,
\end{align}
where $R_A$ ($R_B$) is the subset of sublattice $A$ ($B$) that corresponds to the bit-string $\{r_{2j}\}_{j=1}^N$ ($\{r_{2j-1}\}_{j=1}^N$). Notice that $S_A$ and $R_A$ ($S_B$ and $R_B$) are even-sized subsets of the sublattice $A$ ($B$) of length $N$. Therefore, Eq.~\eqref{FC_O} becomes
\begin{align} \label{FC_O_1D}
    F_O (x,y) = \frac{\sum_{S_A} \left[ \langle \tau_{S_A} \rangle_\beta \langle \tau_{S_A \triangle R_A} \rangle_\beta \right]^{1/2}}{\sum_{S_A} \langle \tau_{S_A} \rangle_\beta} \cdot \frac{\sum_{S_B} \left[ \langle \tau_{S_B} \rangle_\beta \langle \tau_{S_B \triangle R_B} \rangle_\beta \right]^{1/2}}{\sum_{S_B} \langle \tau_{S_B} \rangle_\beta},
\end{align}
which is the product of expressions analogous to Eq.~\eqref{1D_FC_SM}. Such expressions can be computed following the procedure described in Sec.~\ref{Sec:1D_FC:X}. The only difference in the computation is that $\mathbf{r}^{(x,y)}$ needs to be replaced by $\mathbf{r} = \{ r_{2j} \}_{j=1}^N$ (or $\mathbf{r} = \{ r_{2j-1} \}_{j=1}^N$), which yields Eq.~\eqref{1D_FC_Exact} with $|x-y|$ replaced by $|\mathbf{r}| = \sum_{j=1}^N r_{2j}$ (or $|\mathbf{r}| = \sum_{j=1}^N r_{2j-1}$). Since $|\mathbf{r}|$ scales as $|x - y|$, it can similarly be shown as in Eq.~\eqref{FC_exp} that $F_O(x,y)$ decays exponentially for $p < 1/2$. This result demonstrates that SWSSB cannot occur for $p < 1/2$ in the thermodynamic limit.

\subsubsection{Fidelity Correlator for Non-Pauli noise} \label{Sec:1D_FC:NonPauli}

\begin{figure}[t]
    \centering
    \includegraphics[width=0.85\columnwidth]{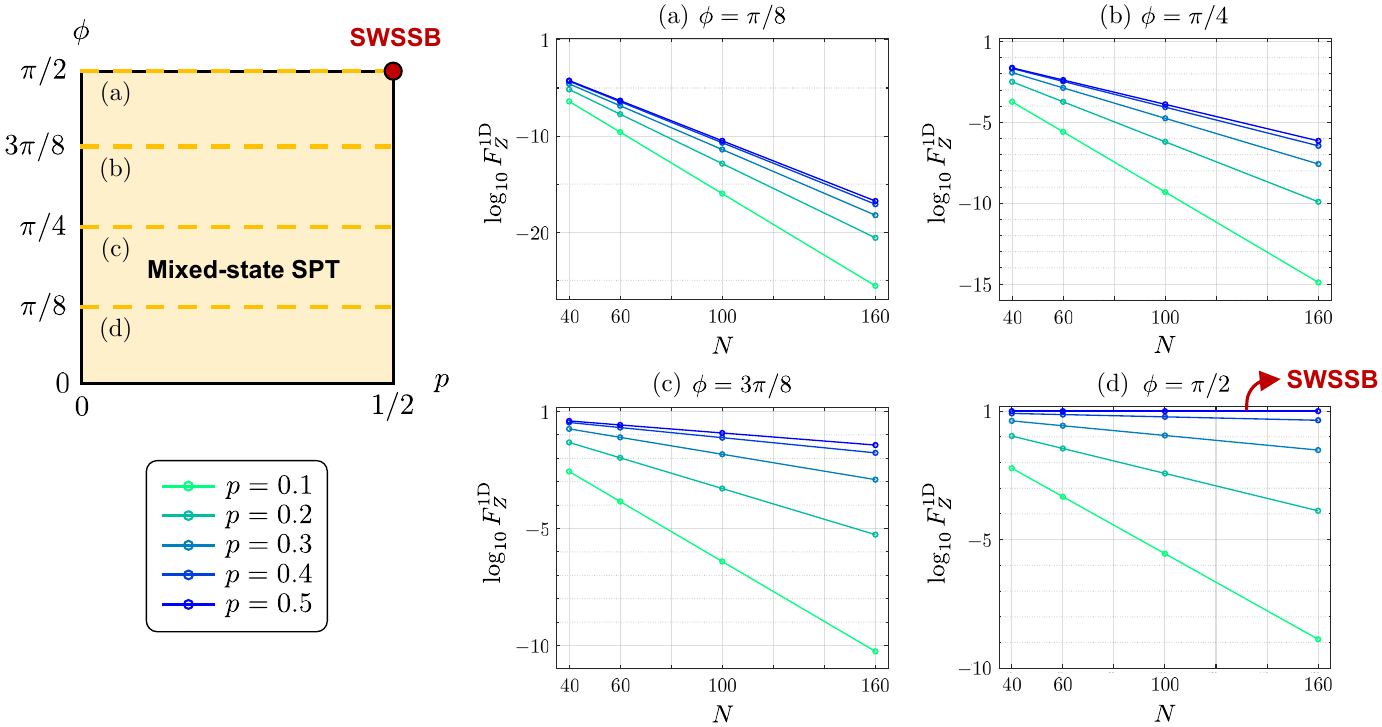}
    \caption{Phase diagram and fidelity correlator $F_Z^{\mathrm{1D}}$ of the 1D cluster state with open boundary conditions under non-Pauli noise $\mathcal{N}^{\mathrm{NP}}$ [see Eq.~\eqref{NP}], shown along four lines: (a) $\phi = \pi/8$, (b) $\phi = \pi/4$, (c) $\phi = 3\pi/8$, and (d) $\phi = \pi/2$.}
    \label{fig:NP}
\end{figure}

Our analytic derivation of maximal robustness applies specifically to generic strongly symmetric local Pauli noise, whose algebraic properties simplify the calculation. Thus, those arguments do not extend directly to generic non-Pauli noises. Nevertheless, we expect the same robustness to hold even for local non-Pauli noises with strong symmetry. To numerically demonstrate this, we consider the following coherent non-Pauli noise $\mathcal{N}^{\mathrm{NP}} = \prod_{j\neq 1, 2N} \mathcal{N}_j^{\mathrm{NP}}$, where
\begin{align} \label{NP}
    \mathcal{N}_j^{\mathrm{NP}} [\rho] = (1-p)\rho + p e^{-i\phi X_j} \rho e^{i\phi X_j},
\end{align}
which applies an $X$-rotation of angle $\phi$ with probability $p$, and respects the strong $\mathbb{Z}_2 \times \mathbb{Z}_2$ symmetry of the cluster state. We adopt open boundary conditions for both numerical convenience and to show that robustness does not rely on the periodic boundary conditions assumed in our analytic treatment. We developed a variational tensor-network method for estimating the fidelity of generic 1D MPOs~\cite{lee2025toappear} and applied it to the open-boundary 1D cluster state under $\mathcal{N}^{\mathrm{NP}}$. The resulting fidelity correlator $F_Z^{\mathrm{1D}}$ and phase diagram are shown in Fig.~\ref{fig:NP}. The fidelity correlator decays exponentially with system size $N$ for all $(p, \phi)$ except the point $(1/2, \pi/2)$, where SWSSB occurs. This confirms that the mixed-state SPT order remains robust against the channel $\mathcal{N}^{\mathrm{NP}}$ in all regimes but $(p, \phi) = (1/2, \pi/2)$. Based on this numerical evidence, we expect similar robustness for generic strongly symmetric local non-Pauli noise. 

\subsection{Entanglement Negativity} \label{Sec:1D_EN}

In Sec.~\ref{Sec:1D_EN:Stab}, the entanglement negativities of $X$- and $Z$-decohered 1D cluster states are computed for $p = 0$ and $1/2$ using the stabilizer formalism. In Sec.~\ref{Sec:1D_EN:General}, the stat-mech expression of the entanglement negativity for general $p$ is discussed.

\subsubsection{Stabilizer Formalism for \texorpdfstring{$p = 0$ and $1/2$}{p = 0 and 1/2}} \label{Sec:1D_EN:Stab}

Let $R$ be a subset of qubits and $\overline{R}$ be its complement. For a stabilizer state $\rho$ with generators $\{ g_1, g_2, \dots, g_{m} \}$, the entanglement negativity $\EE_R = \log \| \rho^{T_R} \|_1$ is given by~\cite{shi2020entanglement,sang2021entanglement}
\begin{align} \label{EN_stab}
    \EE_R = \frac 12 \text{rank} (\mathcal{K}_R) \cdot \log 2,
\end{align}
where $\mathcal{K}_R$ is an $m\times m$ symmetric matrix defined as
\begin{align}
    (\mathcal{K}_R)_{ij} = \begin{cases}
        0 & \text{if } \{ \Pi_R (g_i) , \Pi_R (g_j) \} = 0, \\
        1 & \text{otherwise},
    \end{cases}
\end{align}
and $\Pi_R$ denotes the restriction of stabilizers on the region $R$, i.e., $\Pi_R: g_R \otimes g_{\overline{R}} \mapsto g_R$. Here, $\text{rank}(\mathcal{K}_R)$ is computed using arithmetic modulo 2. For the 1D cluster state, we compute the entanglement negativity with respect to the extensive bipartition $A|B$, where $A$ ($B)$ is the sublattice of odd (even) sites. (This bipartition is referred to as Bravyi's example in Ref.~\cite{zou2016spurious}.)

For $p = 0$ (no decoherence), the density matrix $\rho_0$ is a stabilizer state with stabilizers $K_j = Z_{j-1} X_j Z_{j+1}$. Restricting to sublattice $A$, the stabilizers become $X_j$ for odd $j$ and $Z_{j-1} Z_{j+1}$ for even $j$. The associated matrix $\mathcal{K}_A$ becomes a $2N \times 2N$ circulant adjacency matrix
\begin{align} \label{KA}
    \mathcal{K}_A = \begin{pmatrix}
        0 & 1 & 0 & \cdots & 0 & 1 \\
        1 & 0 & 1 & \cdots & 0 & 0 \\
        0 & 1 & 0 & \cdots & 0 & 0 \\
        \vdots & \vdots & \vdots & \ddots & \vdots & \vdots \\
        0 & 0 & 0 & \cdots & 0 & 1 \\
        1 & 0 & 0 & \cdots & 1 & 0 
    \end{pmatrix},
\end{align}
whose rank is $2N - 2$. Thus, $\EE_A = N \log 2 - \log 2$, and the spurious topological entanglement negativity (TEN) is $\EE_{\text{sp}} = \log 2$.

For $p = 1/2$ (maximal decoherence), note that $\NN_j^P [\rho] = \frac 12 \rho + \frac 12 P_j \rho P_j = \sum_{s_j = \pm 1} M_j^{s_j} \rho M_j^{s_j}$, where $M_j^{s_j} = (1 + s_j P_j) / 2$ is a projector onto the eigenspace of $P_j$ with eigenvalue $s_j$, This is equivalent to measuring $P_j$ without recording the outcome. We can always find generators $\{g_1, \cdots, g_m\}$ of an $n$-qubit stabilizer state $\rho$ such that $\{g_m, P\} = 0$ and $[g_i, P] = 0$ for all $i \neq m$ (by multiplying the other generators that anticommute with $P$ by $g_m$). Then, using $M_j^{s_j} [ (I + g_m)/2 ] M_j^{s_j} = M_j^{s_j} / 2$, we have 
\begin{align}
    \NN_j^P [\rho] = \sum_{s_j = \pm 1}  M_j^{s_j} \left( \prod_{i=1}^{m} \frac{1 + g_i}{2} \right)  M_j^{s_j}= \frac{1}{2^{n-m+1}} \prod_{i=1}^{m-1} \frac{1 + g_i}{2},
\end{align}
i.e., the channel $\NN_j^P$ with $p = 1/2$ removes at most one generator after some redefinition of generators (or none if all generators commute with $P_j$). Consequently, the decohered stabilizer state $\prod_{j=1}^n \NN_j^P [\rho]$ with $p = 1/2$, allowing us to use Eq.~\eqref{EN_stab} to compute its entanglement negativity.

For the 1D cluster state with $p = 1/2$, one can easily see that the surviving generators after applying $\prod_{j=1}^{2N} \NN_j^X$ are the subsystem symmetry generators $G_A = \prod_{j=1}^N X_{2j-1}$ and $G_B = \prod_{j=1}^N X_{2j}$. The restriction to $A$ yields $G_A$ and $I_A$, so $\mathcal{K}_A$ is a zero matrix and hence $\EE_A = 0$ for $\rho_X$. Similarly, no generators survive under $\prod_{j=1}^{2N} \NN_j^Z$, giving $\EE_A = 0$ for $\rho_Z$. Namely, there is no entanglement at all for $\rho_X$ and $\rho_Z$ with $p = 1/2$, which is expected since they are mixtures of classical states obtained by projectively measuring all qubits.

\subsubsection{Entanglement Negativity for General \texorpdfstring{$p$}{p}} \label{Sec:1D_EN:General}

For general $p$, the entanglement negativity of decohered 1D cluster states can be expressed in terms of a stat-mech model. For the sake of clarity, we repeat the derivation for the entanglement negativity of $\rho_X$ given in the main text and then proceed to $\rho_Z$. 

The entanglement negativity of $\rho_X$ is determined by the trace norm $\| (\rho_X)^{T_A} \|_1$, which is the sum of the absolute values of the eigenvalues of $(\rho_X)^{T_A}$. The spectrum of $(\rho_X)^{T_A}$ is given by $(\rho_X)^{T_A} \propto \sum_{\{a\}} \!\left[ \prod_{j=1}^{2N} K_j^{a_j} \right]^{T_A} e^{\beta\sum_{i=1}^{2N} \sigma_i \sigma_{i+2}}$. Note that the Pauli-$Y$ operator flips its sign under transpose. When two adjacent stabilizers $K_i$ and $K_{i+1}$ are present at the same time (i.e., $a_i a_{i+1} = 1$), their product generates two Pauli-$Y$ operators, one in $A$ sublattice and the other in $B$ sublattice. Thus, partial transpose on sublattice $A$ introduces a factor of $-1$ in this case~\cite{lu2020detecting,lu2023characterizing,lu2024disentangling}. This observation leads to $\left[ \prod_{j=1}^{2N} K_j^{a_j} \right]^{T_A} = \prod_{j=1}^{2N} K_j^{a_j} (-1)^{\sum_{i=1}^{2N} a_i a_{i+1}}$, i.e., non-Hermitian nearest-neighbor interactions $(-1)^{a_i a_{i+1}} = \exp \left[\frac{i\pi}{4} (1 - \sigma_i - \sigma_{i+1} + \sigma_i \sigma_{i+1}) \right]$ among the Ising spins $\sigma_i = 1 - 2a_j \in \{\pm 1\}$ are generated along the entangling boundary due to partial transpose. Using the relation $K_j^{a_j} = \sigma_j^{(1 -K_j)/2}$, we obtain Eq.~\eqref{1D_TraceNorm} of the main text:
\begin{align} 
    \| (\rho_X)^{T_A} \|_1 = \frac{1}{C_X} \sum_S \left| \left\langle \prod_{i\in S} \sigma_i \right\rangle_{\!\beta,X} \right|,
\end{align}
where $S$ runs over all subsets of the length-$2N$ chain and $\langle \cdot \rangle_{\beta,X}$ represents the expectation value in the non-Hermitian 1D stat-mech model $H_{\beta,X} = \sum_{i=1}^{2N} \left( \frac{i\pi}{4} \sigma_i \sigma_{i+1} + \beta \sigma_i \sigma_{i+2} - \frac{i\pi}{2} \sigma_i \right)$. Here, $C_X$ is a normalization constant:
\begin{align}
    C_X = \sum_S \left\langle \prod_{i\in S} \sigma_i \right\rangle_{\!\beta,X} = \left\langle \prod_{j=1}^{2N} (1 + \sigma_j) \right\rangle_{\!\beta,X} = 2^{2N} \left\langle \prod_{j=1}^{2N} \delta_{\sigma_j, 1} \right\rangle_{\!\beta,X}.
\end{align}
Note that $\left\langle \prod_{j=1}^{2N} \delta_{\sigma_j, 1} \right\rangle_{\!\beta,X} = (T_{\beta,X})_{11}^{2N} / Z_{\beta,X}$, where 
\begin{align}
    T_{\!\beta,X} = \begin{pmatrix}
        e^{\beta - i\pi/4} & e^{-\beta - i\pi/2} & 0 & 0 \\
        0 & 0 & e^{\beta + i\pi/4} & e^{-\beta + i\pi/2} \\
        e^{-\beta - i\pi/2} & e^{\beta - 3i\pi/4} & 0 & 0 \\
        0 & 0 & e^{-\beta + i\pi/2} & e^{\beta + 3i\pi/4}
    \end{pmatrix}
\end{align}
is the transfer matrix of the model $H_{\beta,X}$ written in the ordered basis $\{ |{+}1,{+}1\rangle, |{+}1,{-}1\rangle, |{-}1,{+}1\rangle, |{-}1,{-}1\rangle \}$ and
\begin{align}
    Z_{\beta, X} = \Tr [T_{\beta,X}^{2N}] = \frac{2 \big[ (\sqrt{2e^{4\beta} - 1} + 1)^{2N} + (\sqrt{2e^{4\beta} - 1} - 1)^{2N} \big]}{(2i)^N e^{2\beta N}}
\end{align}
is the partition function of $H_{\beta, X}$. Thus, the constant $C_X$ simplifies to
\begin{align}
    C_X = \frac{2^{3N-1} e^{4\beta N}}{(\sqrt{2e^{4\beta} - 1} + 1)^{2N} + (\sqrt{2e^{4\beta} - 1} - 1)^{2N}}.
\end{align}

Similarly, we can express $\| \rho_Z \|_1$ in terms of a non-Hermitian stat-mech model. Under conjugation by $Z_i$, $\prod_{j=1}^{2N} K_j^{a_j}$ flips (maintains) its sign when $\sigma_i = -1$ ($+1$). Thus, there are only Zeeman terms in the spectrum of $\rho_Z$:
\begin{align}
    \rho_Z \propto \sum_{\{a\}} \prod_{j=1}^{2N} K_j^{a_j} (1-2p)^{\sum_{i=1}^{2N} \frac{1 - \sigma_i}{2}} \propto \sum_{\{a\}} \prod_{j=1}^{2N} K_j^{a_j} e^{\beta \sum_{i=1}^N \sigma_i},
\end{align}
with $\beta = -\frac 12 \log(1-2p)$. Now, using the relation $\left[ \prod_{j=1}^{2N} K_j^{a_j} \right]^{T_A} = \prod_{j=1}^{2N} K_j^{a_j} (-1)^{\sum_{i=1}^{2N} a_i a_{i+1}}$ again, we obtain
\begin{align} 
    \| (\rho_Z)^{T_A} \|_1 = \frac{1}{C_Z} \sum_S \left| \left\langle \prod_{i\in S} \sigma_i \right\rangle_{\!\beta,Z} \right|,
\end{align}
where $S$ runs over all subsets of the length-$2N$ chain, $\langle \cdot \rangle_{\beta,Z}$ is the expectation value in the non-Hermitian 1D stat-mech model $H_{\beta,Z} = \sum_{i=1}^{2N} \left[ \frac{i\pi}{4} \sigma_i \sigma_{i+1} + \left( \beta - \frac{i\pi}{2} \right) \sigma_i \right]$, whose transfer matrix is 
\begin{align}
    T_{\beta,Z} = \begin{pmatrix}
        e^\beta & 1 \\
        1 & -e^{-\beta}
    \end{pmatrix}
\end{align}
(up to an unimportant phase factor) and the partition function is
\begin{align}
    Z_{\beta,Z} = \Tr [T_{\beta,Z}^{2N}] = \frac{(e^{2\beta} - 1 + \sqrt{e^{4\beta} + 6e^{2\beta} + 1})^{2N} + (e^{2\beta} - 1 - \sqrt{e^{4\beta} + 6e^{2\beta} + 1})^{2N}}{2^{2N} e^{2\beta N}}.
\end{align}
The normalization constant $C_Z$ is given by
\begin{align}
    C_Z = \sum_S \left\langle \prod_{i\in S} \sigma_i \right\rangle_{\!\beta,Z} = \frac{2^{2N} (T_{\beta,Z})_{11}^{2N}}{Z_{\beta,Z}} = \frac{2^{4N} e^{4\beta N}}{(e^{2\beta} - 1 + \sqrt{e^{4\beta} + 6e^{2\beta} + 1})^{2N} + (e^{2\beta} - 1 - \sqrt{e^{4\beta} + 6e^{2\beta} + 1})^{2N}}.
\end{align}

\subsection{Proof of Theorem} \label{Sec:1D_Thm1}

In Sec.~\ref{Sec:Thm1_Pre}, we introduce terminologies and review key preliminary concepts before presenting the proof of Theorem in Sec.~\ref{Sec:Thm1_Proof}.

\subsubsection{Preliminaries} \label{Sec:Thm1_Pre}

\noindent \textbf{(1) R\'enyi Entanglement Negativity} \vspace{2pt} 

For a given density matrix $\rho$, the (logarithmic) entanglement negativity of a subsystem $R$ is defined as
\begin{align}
    \EE_R = \log \| \rho^{T_R} \|_1,
\end{align}
where $T_R$ denotes the partial transpose on $R$ and $\| \cdot \|_1$ is the trace norm. Entanglement negativity is an easily computable entanglement measure for mixed states and upper-bounds distillable entanglement. However, computing $\|\rho^{T_R} \|_1$ for quantum many-body states is challenging, as it requires the full spectrum of $\rho^{T_R}$. A more tractable and commonly used alternative is the \emph{R\'enyi-$(2\alpha)$ entanglement negativity}, defined as
\begin{align} \label{RenyiEN}
    \EE_R^{(2\alpha)} = \frac{1}{2 - 2\alpha} \log \left( \frac{\Tr [(\rho^{T_R})^{2\alpha}]}{\Tr [\rho^{2\alpha}]} \right).
\end{align}
Although $\EE_R^{(2\alpha)}$ is not strictly an entanglement monotone, it qualitatively captures the behavior of $\EE_R$ in quantum many-body states and reduces to $\EE_R$ in the limit $\alpha \rightarrow 1/2$. For integer $\alpha$, $\EE_R^{(2\alpha)}$ involves the $(2\alpha)$th moments of $\rho$ and $\rho^{T_R}$, which depend on $2\alpha$ copies of these matrices. This allows for analytic computations using techniques such as the replica trick~\cite{calabrese2012entanglement}. \vspace{7pt}

\noindent \textbf{(2) MPS, Injectivity and 1D SPT Order} \vspace{2pt}

A 1D short-range entangled quantum state can be efficiently represented by a \emph{matrix product state (MPS)} with a bond dimension $D$ independent of the system size~\cite{schollwock2011density}. Consider a translationally invariant MPS 
\begin{align}
    |\psi\rangle = \sum_{i_1, \dots, i_N = 1}^d \Tr [M^{i_1} \cdots M^{i_N}] | i_1, \dots, i_N \rangle,
\end{align}
where $d$ is the physical dimension and $M^i$ are $D\times D$ matrices. The MPS is \emph{injective} if the MPS tensor $M: \C^{D^2} \rightarrow \C^d$ defines an injective map from the virtual space $\C^{D^2}$ to the physical space $\C^d$~\cite{perezgarcia2007matrix}. The injectivity of the MPS, which holds for generic MPSs, ensures short-range correlation. It also guarantees that the transfer matrix $T = \sum_{i=1}^d M^i \otimes (M^i)^*$ has a unique largest real eigenvalue and the map $X \mapsto \sum_i M^i X (M_i)^\dagger$ is a full-rank linear map on the $D^2$-dimensional space of matrices.

For an injective MPS with an on-site $G$ symmetry represented by $U_{g} = \prod_{j=1}^N u_g$ ($g \in G$), there exist $D\times D$ invertible matrices $V_g$ satisfying~\cite{perezgarcia2008string}
\begin{align} \label{MPS_sym}
    \sum_{j_1 = 1}^d (u_g)_{ij} M^j = e^{i\theta_g} V_g^{-1} M^i V_g^,
\end{align}
where $\theta_g$ is a phase. Graphically, symmetry fractionalization in Eq.~\eqref{MPS_sym} can be depicted as
\begin{equation}
    \begin{aligned}
        \includegraphics[height=1.6cm]{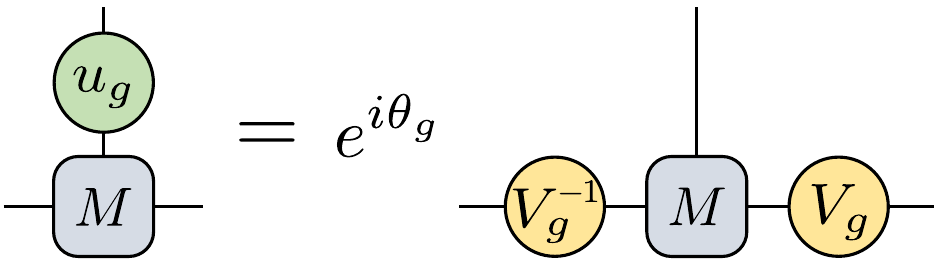}.
    \end{aligned}
\end{equation}
Here, the map $V: g \mapsto V_g$ forms a projective representation of $G$, i.e., $V_{g_1} V_{g_2} = \omega(g_1, g_2) V_{g_1 g_2}$ for all $g_{1,2} \in G$, where the phase factor $\omega: G \times G \rightarrow \mathrm{U}(1)$ satisfying $\omega (g_1, g_2) \omega (g_1 g_2, g_3) = \omega(g_1, g_2 g_3) \omega(g_2, g_3)$ for all $g_{1,2,3} \in G$ is called a \emph{2-cocycle} (or a \emph{factor system}). Redefining $V'_g = \alpha_g V_g$ with $\alpha_g \in \mathrm{U}(1)$, the 2-cocycle transforms as $\omega'(g_1, g_2) = (\alpha_{g_1 g_2} / \alpha_{g_1} \alpha_{g_2}) \cdot \omega (g_1, g_2)$. This relation between the 2-cocycles $\omega(g_1, g_2)$ and $\omega'(g_1,g_2)$ defines the equivalence class $[\omega]$ of projective representations, which form the \emph{second cohomology group} $\mathcal{H}^2 (G, \mathrm{U}(1))$~\cite{chen2011classification}. The identity element of $\mathcal{H}^2 (G, \mathrm{U}(1))$ represents linear representations, whereas the remaining elements correspond to nontrivial projective representations.

When $|\psi \rangle$ exhibits a nontrivial SPT order, the projective representation $V$ in Eq.~\eqref{MPS_sym} belongs to a nontrivial 2-cocycle class. In particular, for a finite Abelian group $G$, we have $V_{g_1} V_{g_2} = \phi (g_1, g_2) V_{g_2} V_{g_1}$, where the \emph{slant product} $\phi(g_1, g_2) = \omega(g_1, g_2) / \omega(g_2, g_1)$ characterizes the equivalence class of $V$. In this case, the values of $\phi (g_1, g_2)$ are always roots of unity, i.e., $\phi (g_1, g_2) = e^{2\pi i p / q}$ where $p$ and $q > 1$ are coprime integers. The integer $q$, called the \emph{order} of the 2-cocycle class, is determined by the structure of $\mathcal{H}^2 (G, \mathrm{U}(1))$.

For example, the 1D cluster state represented by the MPS tensors 
\begin{align}
    C^1 = \begin{pmatrix}
        1 & 1 \\ 0 & 0
    \end{pmatrix}, \qquad
    C^2 = \begin{pmatrix}
        0 & 0 \\ 1 & -1
    \end{pmatrix}
\end{align}
has a $\Z_2 \times \Z_2$ symmetry generated by $G_A = \prod_{j:\mathrm{odd}} X_j$ and $G_B = \prod_{j:\mathrm{even}} X_j$. Grouping two neighboring sites, one can directly verify Eq.~\eqref{MPS_sym}: 
\begin{equation}
    \begin{aligned}
        \includegraphics[height=1.6cm]{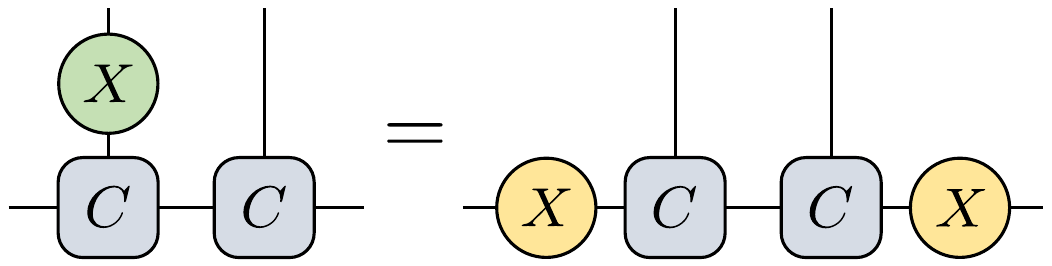}, \qquad \includegraphics[height=1.6cm]{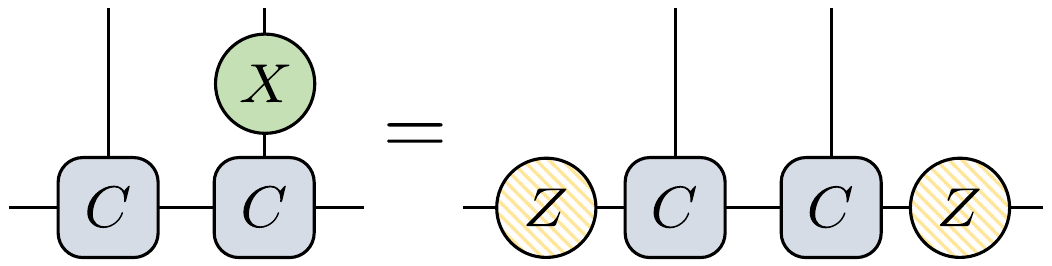}.
    \end{aligned}
\end{equation}
Since $XZ = -ZX$, the 1D cluster state hosts a nontrivial $\Z_2 \times \Z_2$ SPT order with $q = 2$. \vspace{7pt}

\noindent \textbf{(3) MPDO, Strong Injectivity, Non-Degenerate Channels and 1D Mixed-State SPT Order} \vspace{2pt}

This part mainly summarizes key concepts from Ref.~\cite{xue2024tensor}. A 1D mixed state can be represented by a \emph{matrix product density operator (MPDO)}~\cite{verstraete2004matrix}. A translationally invariant MDPO takes the form
\begin{align} \label{MPDO}
    \rho = \sum_{\substack{i_1, \dots, i_N = 1\\j_1, \dots, j_N = 1}}^d \Tr[M^{i_1 j_1} \cdots M^{i_1 j_1}] | i_1, \dots, i_N \rangle \langle j_1, \dots, j_N |,
\end{align}
where $M^{ij}$ are $D\times D$ matrices. To represent a physical density matrix, the MPDO must satisfy (i) Hermiticity: $\rho = \rho^\dagger$, (ii) positive semidefiniteness: $\rho \geq 0$, and (iii) the unit-trace condition: $\Tr [\rho] = 1$. The MPDO in is \emph{strongly injective} if it satisfies the following two properties~\cite{xue2024tensor,guo2024locally}:
\begin{enumerate}
    \item[(C1)] The map $M: \C^{D^2} \rightarrow \C^{d^2}$ is an injective map from the virtual space $\C^{D^2}$ to the physical space $\C^{d^2}$.

    \item[(C2)] The transfer matrix $\TT \equiv \sum_{i=1}^d M^{ii}$ has a unique largest real eigenvalue.
\end{enumerate}
Conditions (C1) and (C2) ensure short-range correlations in the doubled state $| \rho\rangle\!\rangle$ and the density matrix $\rho$, respectively. Strong injectivity of MPDO has been a crucial technical condition for classifying 1D mixed-state SPT orders~\cite{xue2024tensor,guo2024locally}. 

We introduce an additional technical condition that generalizes (C1). Define the MPDO tensor $M_\alpha$ by contracting $\alpha \in \N$ copies of the given MPDO tensor $M$ along the physical space direction as follows:
\begin{equation} \label{M_alpha}
    \begin{aligned}
        \includegraphics[height=2.4cm]{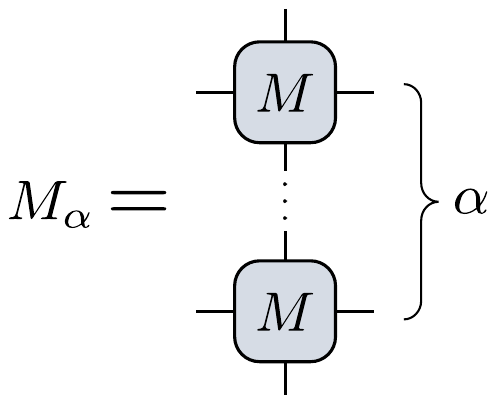}.
    \end{aligned}
\end{equation}
The generalized condition goes as follows:
\begin{enumerate}
    \item[(C1')] For all $\alpha \in \N$, $M_\alpha: \C^{D^{2\alpha}} \rightarrow \C^{d^2}$ is an injective map from the virtual space $\C^{D^{2\alpha}}$ to the physical space $\C^{d^2}$.
\end{enumerate}
Condition (C1') holds for generic MPDOs. [$\because$ The violation of condition (C1') for some $\alpha$ is equivalent to the existence of matrix $X_\alpha \neq 0$ such that $\sum_{i,j} M_\alpha^{ij} X_\alpha (M_\alpha^{ij})^\dagger = 0$. Each such vanishing condition is a polynomial equation in the entries of $M$, so its solution set has measure zero. Since a countable union of measure-zero sets remains measure zero, MPDO tensors satisfy condition (C1)' for all $\alpha \in \N$ with probability 1.] Physically, condition (C1') ensures short-range correlations in the doubled state $| \rho^\alpha \rangle\!\rangle$.

An MPDO admits a \emph{local purification} $| \Psi \rangle_{SE}$ if it arises from tracing out the environment $E$ from a pure state $| \Psi \rangle_{SE}$, i.e., $\rho_S = \Tr_E [| \Psi \rangle \langle \Psi|_{SE}]$. This includes cases where a pure state undergoes decoherence via a local quantum channel. If the local purification has an MPS representation 
\begin{align}
    |\Psi\rangle_{SE} = \sum_{i_1, \dots, i_N = 1}^d \sum_{k_1, \dots, k_N} \Tr[\tilde{M}^{i_1 k_1} \cdots \tilde{M}^{i_N k_N}] | i_1, \dots, i_N \rangle_S |k_1, \dots, k_N \rangle_E,
\end{align}
the corresponding MPDO tensor satisfies $M^{ij} = \sum_k \tilde{M}^{ik} \otimes (\tilde{M}^{jk})^*$. If condition (C1) holds for an MPDO $\rho_S$ with a local purification $|\Psi\rangle_{SE}$, then condition (C2) automatically follows. This follows from the fact that the MPDO transfer matrix $\TT = \sum_{i=1} ^d \sum_k \tilde{M}^{ik} \otimes (\tilde{M}^{ik})^*$ equals the transfer matrix of the purification $|\Psi\rangle_{SE}$, which has a unique largest real eigenvalue since condition (C1) is equivalent to the injectivity of $| \Psi \rangle_{SE}$.

Next, a quantum channel $\NN[\cdot]$ is called \emph{non-degenerate} if it is an injective map acting on the operator Hilbert space $\C^{d^2}$. Such channels describe Lindbladian evolutions over finite time. Given a Kraus representation $\NN[\cdot] = \sum_l K_l [\cdot] K_l^\dagger$, non-degeneracy of $\NN$ requires the matrix $\sum_l K_l \otimes K_l^*$ to have a full rank. It is shown in Ref.~\cite{xue2024tensor} that $\rho$ is strongly injective if and only if $\NN[\rho]$ is strongly injective, where $\NN$ is a finite-depth brickwork circuit composed of local non-degenerate channels. This implies that an injective pure state (which is trivially strongly injective) retains strong injectivity after decoherence by non-degenerate channels. 

For example, consider an incoherent Pauli noise $\NN^P [\cdot] = (1-p) [\cdot] + p P [\cdot] P$, where $P$ is a Pauli-string operator and $p \in [0,1/2]$. Since $P^2 = I$, the associated matrix $(1-p) I\otimes I + p P \otimes P^*$ has eigenvalues $1$ and $1-2p$. This shows that $\NN^P$ is non-degenerate for all $p \in [0, 1/2)$, in which case neither eigenvalue is zero. Namely, $\NN^P$ are not non-degenerate only at the maximal error rate $p = 1/2$. This aligns with its representation as a Lindbladian evloution $e^{t \mathcal{L}}$ with $\mathcal{L} [\cdot] = \frac 12 (P [\cdot] P - [\cdot])$ and $t = -\log (1-2p)$, where the evolution time $t$ diverges at $p = 1/2$. 

Finally, we discuss the classification of 1D mixed-state SPT order. The SPT order of a strongly injective MPDO $\rho$ is defined by the SPT order of the corresponding doubled state $|\rho\rangle\!\rangle$, whose MPS representation is injective~\cite{xue2024tensor,guo2024locally}. Consequently, the classification scheme for MPS discussed in Sec.~\ref{Sec:Thm1_Pre}(b) can be directly applied to MPDOs. A strong symmetry $G$ of an MPDO $\rho$ translates into a $G^u \times G^l$ symmetry of the doubled state, where $G^u$ ($G^l$) represents a symmetry $G$ acting on the ket (bra) Hilbert space. Under the symmetry $(G^u \times G^l) \times K$ with $K$ being a possible weak symmetry, the classification of 1D mixed-state SPT phases is given by the cohomology group $\mathcal{H}^2 (G, \mathrm{U}(1)) \oplus \mathcal{H}^1 (G, \mathcal{H}^1 (K, \mathrm{U}(1))$~\cite{ma2023average,xue2024tensor,guo2024locally}. Again, the nontrivial mixed-state SPT order means the corresponding projective representations are nontrivial. 

For an MPDO $\rho$ with only strong symmetry $G$, the classification matches the pure-state case, given by $\mathcal{H}^2 (G, \mathrm{U}(1))$. The symmetry fractionalization follows a form similar to Eq.~\eqref{MPS_sym}:
\begin{equation} \label{MPDO_Sym}
    \begin{aligned}
        \includegraphics[height=1.73cm]{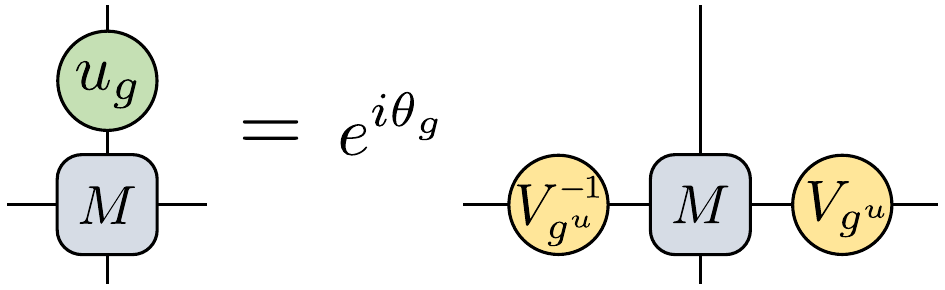}, \qquad
        \includegraphics[height=1.73cm]{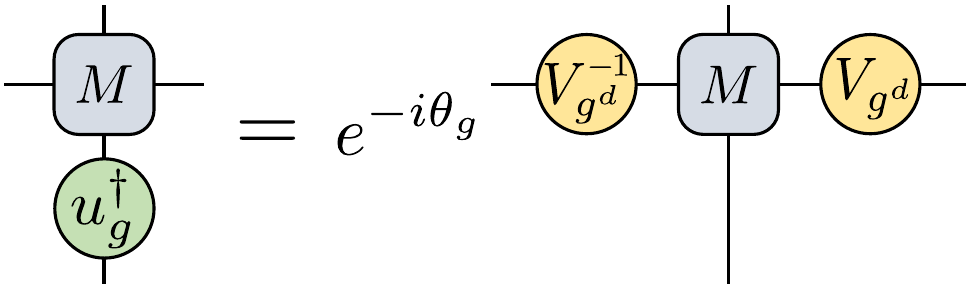}. 
    \end{aligned}
\end{equation}
Here, the Hermiticity constraint is used to relate phase factors. The Hermiticity can also be employed to impose $\phi(g^u, h^u) = \phi^* (g^d, h^d)$ for all $g^{u/d}, h^{u/d} \in G^{u/d}$~\cite{xue2024tensor}. Additionally, it is known that no mixed anomaly exists between the ket and bra symmetries $G^{u,d}$, implying $[V_{g^u}, V_{g^d}] = 0$ for all $g^{u/d} \in G^{u/d}$~\cite{xue2024tensor,duivenvoorden2017entanglement}. 

\subsubsection{Proof of Theorem} \label{Sec:Thm1_Proof}

Here, we prove Theorem of the main text. Let's restate it for convenience:
\begin{theorem*}
Consider a 1D nontrivial mixed-state SPT state $\rho$ protected by on-site $G_1 \times G_2$ symmetry, where $G_i$ are finite Abelian groups acting on separate bipartite subsystems, and let $q$ be the order of the 2-cocycle class associated with the SPT order. Then, $\rho$ has a R\'enyi spurious TEN $\EE_{\text{sp}}^{(2\alpha)} \geq \log q$ for all $\alpha \in \Z_{\geq 2}$ with respect to the bipartition, provided that the MPDO representation of $\rho$ is strongly injective and satisfies condition (C1').
\end{theorem*}

\begin{figure}[b]
    \includegraphics[width=0.79\columnwidth]{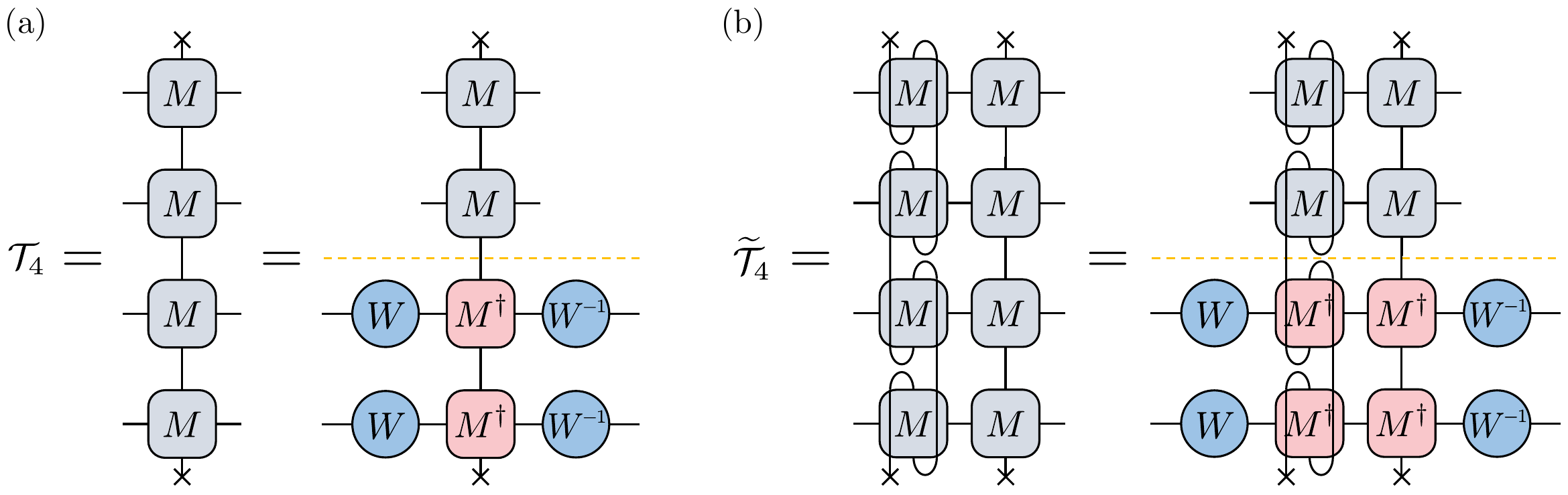}
    \caption{(a) A transfer matrix $\TT_{2\alpha}$ is related to the MPS transfer matrix of $M_\alpha$ via a similarity transformation. (b)
    Similarly, $\widetilde{\TT}_{2\alpha}$ is also related to some completely positive map via a similarity transformation. The case with $\alpha = 2$ is shown for simplicity. The cross marks ($\times$) indicate contractions along the vertical directions. }
    \label{fig:lem1}
\end{figure}

\noindent \emph{Proof of Theorem.} Consider a strongly injective MPDO, which can be graphically represented as follows:
\begin{equation}
    \begin{aligned}
        \includegraphics[height=1.1cm]{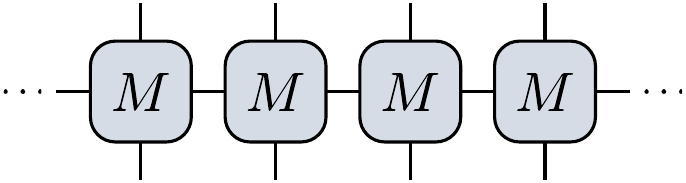}.
    \end{aligned}
\end{equation}
With this MPDO, the moments appearing in the R\'enyi-$(2\alpha)$ negativity $\EE_A^{(2\alpha)}$ for $\alpha \in \Z_{\geq 2}$ [see Eq.~\eqref{RenyiEN}] can be expressed as
$\Tr[\rho^{2\alpha}] = \Tr[\TT_{2\alpha}^{2N}]$ and $\Tr[(\rho^{T_A})^{2\alpha}] = \Tr[\widetilde{\TT}_{2\alpha}^N]$, where the transfer matrices $\TT_{2\alpha}$ and $\widetilde{\TT}_{2\alpha}$ are defined as
\begin{equation}
    \begin{aligned}
        \includegraphics[height=4.7cm]{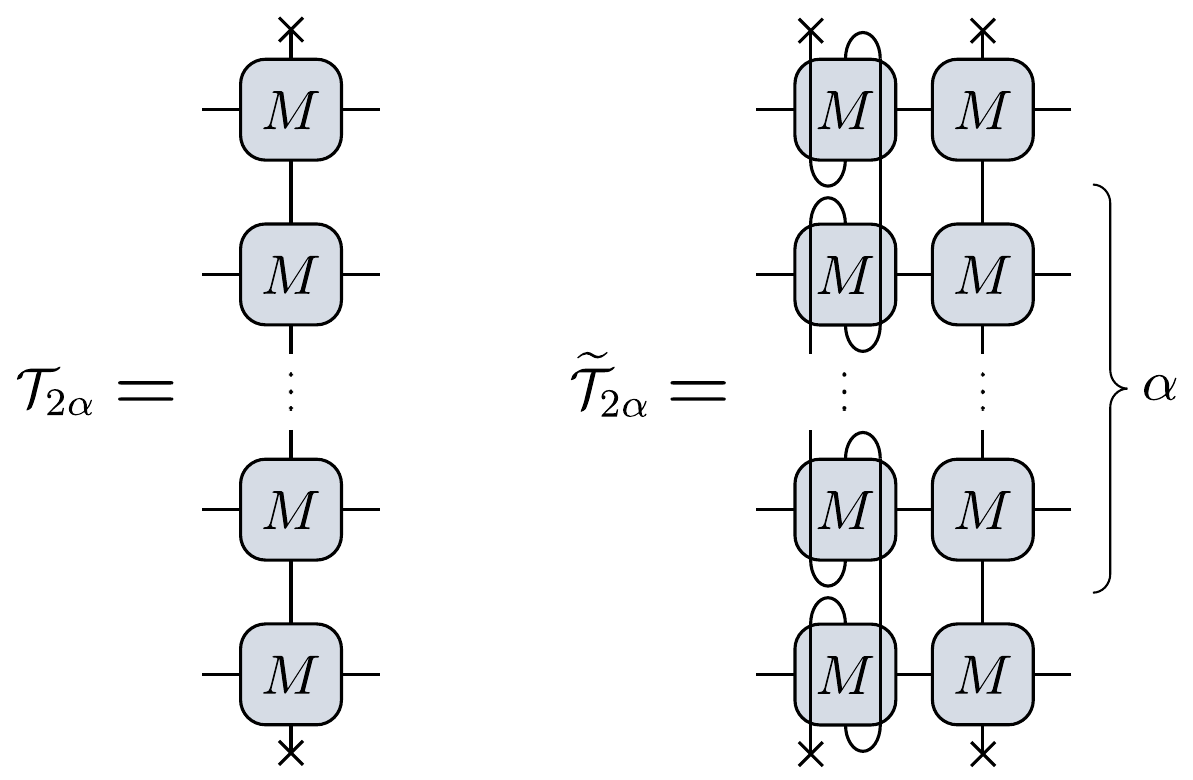}
    \end{aligned}
\end{equation}
and act on a virtual space of dimension $D^{2\alpha}$. (The system's linear size is given by $2N$.) We begin by proving a lemma on the spectrum of the transfer matrices:
\begin{lemma*}
For a strongly injective MPDO, all eigenvalues of the transfer matrices $\TT_{2\alpha}$ and $\widetilde{\TT}_{2\alpha}$ with the largest modulus are real for all $\alpha \in \N$. Assuming condition (C1'), $\TT_{2\alpha}$ has a unique largest real eigenvalue for all $\alpha \in \N$.
\end{lemma*}
\emph{Proof.} The Hermiticity condition ($\rho^\dagger = \rho$) and condition (C1) imply that $(M^{ji})^* = W^{-1} M^{ij} W$ for some invertible matrix $W$ (a possible phase factor can be absorbed into $M^{ij}$). Then, the MPDO $M_{2\alpha}$ [see Eq.~\eqref{M_alpha}] admits a local purification with an MPS tensor given by $M_\alpha$, up to a similarity transformation acting on the half of the virtual space [see Fig.~\ref{fig:lem1}(a) for the case $\alpha = 2$]. Thus, the MPDO transfer matrix of $M_{2\alpha}$, which is nothing but $\TT_{2\alpha} = \sum_{i = 1}^d M_{2\alpha}^{ii}$, is related via a similarity transformation to the transfer matrix of the MPS $M_\alpha$, which is completely positive. Since the largest-modulus eigenvalues of completely positive maps are always real, $\TT_{2\alpha}$ shares this spectral property. One can similarly relate $\widetilde{\TT}_{2\alpha}$ with a completely positive map via a similarity transformation as shown in Fig.~\ref{fig:lem1}(b) and yield the same conclusion.

Under an additional condition (C1'), the map $M_\alpha: \C^{D^{2\alpha}} \rightarrow \C^{d^2}$ is an injective map for all $\alpha \in \N$. This means the MPS transfer matrix of $M_\alpha$ has a unique largest real eigenvalue for all $\alpha \in \N$. Since $\TT_{2\alpha}$ shares the same spectrum with the MPS transfer matrix of $M_\alpha$, the proof is complete. \hfill $\square$
\vspace{4pt}

We now apply an argument similar to that in Ref.~\cite{zou2016spurious} to the transfer matrix $\widetilde{T}_{2\alpha}$ to prove Theorem. As discussed in Sec.~\ref{Sec:Thm1_Pre}(3), an MPDO with a nontrivial mixed-state SPT order protected by $G_1 \times G_2$ symmetry transforms as follows:
\begin{equation} \label{Thm1_Sym1}
    \begin{aligned}
        \includegraphics[height=1.73cm]{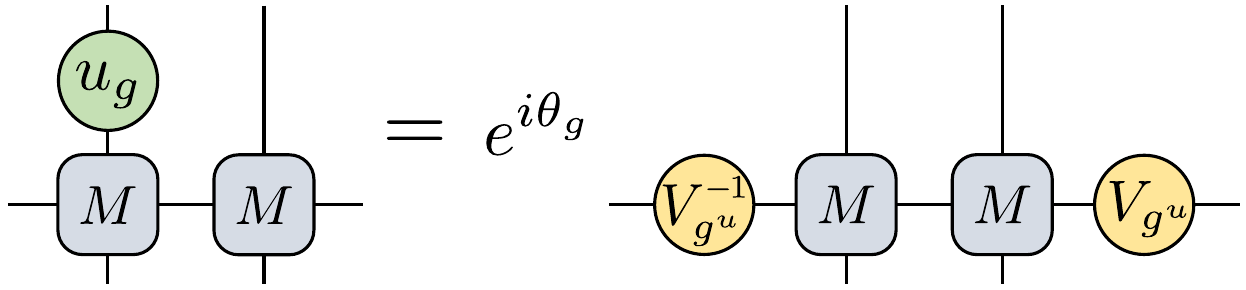}, \qquad
        \includegraphics[height=1.73cm]{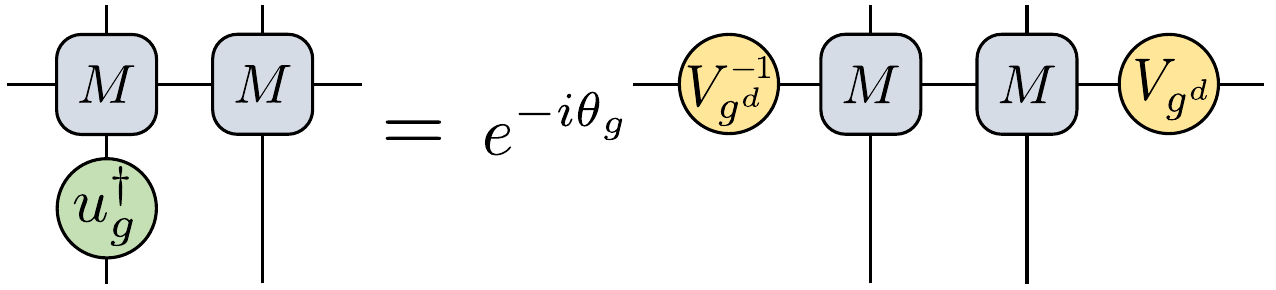}, \\
        \includegraphics[height=1.73cm]{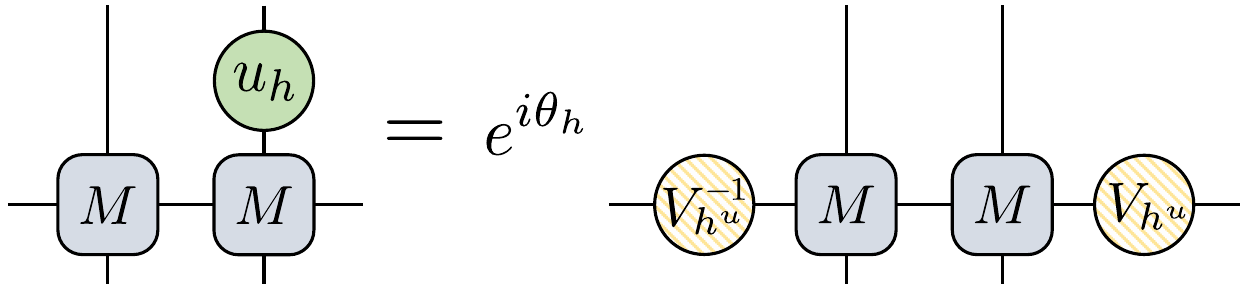}, \qquad
        \includegraphics[height=1.73cm]{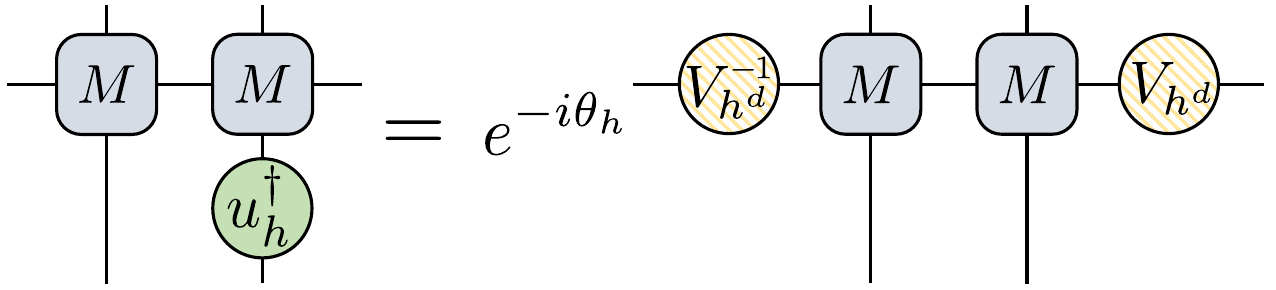},
    \end{aligned}
\end{equation}
where $V_{g^u} V_{h^u} = \Omega V_{h^u} V_{g^u}$ and $V_{g^d} V_{h^d} = \Omega^{-1} V_{h^d} V_{g^d}$, with $\Omega = e^{2\pi i p/q}$ for coprime integers $p$ and $q > 1$. Additionally, the projective representations $V$ of $G^u$ commute with those of $G^d$. 

\begin{figure}[t]
    \includegraphics[width=0.94\columnwidth]{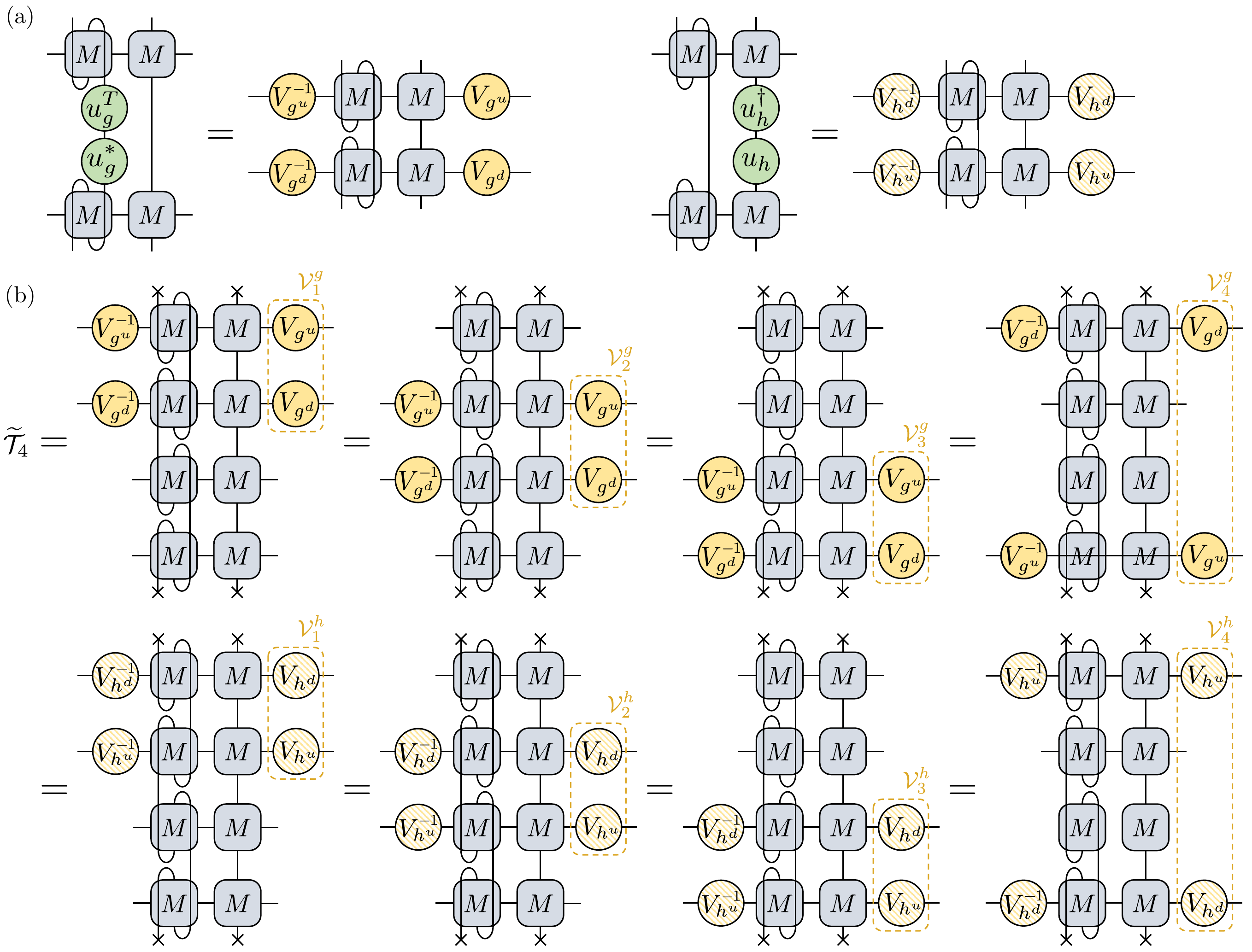}
    \caption{(a) Identities that follow from the MPDO symmetry transformation rule in Eq.~\eqref{MPDO_Sym}. (b) The transfer matrix $\widetilde{\TT}_{2\alpha}$ remains invariant under the similarity transformations induced by $\VV_j^g$ ($\VV_j^h$) that follow from the right (left) identity of (a) Here, the case $\alpha = 2$ is illustrated.}
    \label{fig:thm1}
\end{figure}

From Eq.~\eqref{Thm1_Sym1}, the identities in Fig.~\ref{fig:thm1}(a) follow directly. Using these identities, one can see that $\widetilde{\TT}_{2\alpha}$ is invariant under a series of similarity transformations, as shown in Fig.~\ref{fig:thm1}(b) for the case $\alpha = 2$. These transformations are induced by the matrices $\VV_j^g = V_{g^u}^{(j)} \otimes V_{g^d}^{(j+1)}$ and $\VV_j^h = V_{h^d}^{(j)} \otimes V_{h^u}^{(j+1)}$, where superscripts indicate legs of the virtual space from top to bottom, with $2\alpha + 1 \equiv 1$. These matrices satisfy the following algebra:
\begin{align} \label{alg}
    \VV_j^g \VV_{j+1}^h = \Omega^{-1} \VV_{j+1}^h \VV_j^g, \qquad \VV_{j+1}^g \VV_j^h = \Omega \VV_j^h \VV_{j+1}^g, \qquad 1\leq j\leq 2\alpha,
\end{align}
while all other matrices commute. This is a direct sum of two independent algebras of the same structure:
\begin{align} \label{algalg}
    \begin{cases}
        \VV_{2j-1}^g \VV_{2j}^h = \Omega^{-1} \VV_{2j}^h \VV_{2j-1}^g, \\[3pt]
        \VV_{2j+1}^g \VV_{2j}^h = \Omega \VV_{2j}^h \VV_{2j+1}^g,
    \end{cases} \qquad \text{and} \qquad \begin{cases}
        \VV_{2j}^g \VV_{2j+1}^h = \Omega^{-1} \VV_{2j+1}^h \VV_{2j}^g, \\[3pt]
        \VV_{2j}^g \VV_{2j-1}^h = \Omega \VV_{2j-1}^h \VV_{2j}^g,
    \end{cases}\quad  1\leq j\leq \alpha,
\end{align}
Focusing on the right algebra of Eq.~\eqref{algalg} (the left follows similarly), note that the same algebra can be generated by $\mathcal{Z}_j = \VV_{2j}^h$ and $\mathcal{X}_j = \prod_{k=1}^j \VV_{2k-1}^g$ for $1\leq j\leq \alpha - 1$ because $\prod_{j=1}^\alpha \VV_{2j-1}^g$ and $\prod_{j=1}^\alpha \VV_{2j}^h$ are in the center of the algebra. Since these new generators satisfy $\mathcal{Z}_j \mathcal{X}_j = \Omega \mathcal{X}_j \mathcal{Z}_j$ ($1\leq j\leq \alpha - 1$) with all other generators commuting, the minimal representation of the right algebra has dimension $q^{\alpha - 1}$. Therefore, the dimension of the minimal representation of the full algebra in Eq.~\eqref{alg} is $q^{2(\alpha - 1)}$.

Finally, let $\lambda_1$ ($\widetilde{\lambda}_1$) be the eigenvalues of $\TT_{2\alpha}$ ($\widetilde{\TT}_{2\alpha}$) with the largest modulus and $d_1$ ($\widetilde{d}_1$) their degeneracy. Since $\lambda_1$, $\widetilde{\lambda}_1$ are real and $d_1 = 1$ by Lemma, the R\'enyi-($2\alpha$) negativity can be written as
\begin{align}
    \EE_A^{(2\alpha)} = \frac{1}{2 - 2\alpha} \log \left( \frac{\Tr [\widetilde{\TT}_{2\alpha}^N]}{\Tr [\TT_{2\alpha}^{2N}]} \right) \underset{N \rightarrow \infty}{\longrightarrow} \frac{\log(\lambda_1^2 / \widetilde{\lambda}_1)}{2(\alpha-1)} N - \underbrace{\frac{ \log (\widetilde{d}_1)}{2(\alpha-1)}}_{=\, \EE_{\mathrm{sp}}^{(2\alpha)}},
\end{align}
where the first term describes area-law scaling, and the second is the R\'enyi-($2\alpha$) spurious TEN $\EE_{\mathrm{sp}}^{(2\alpha)}$. Since the algebra Eq.~\eqref{alg} at least requires a dimension of $q^{2(\alpha)}$, it follows that $\widetilde{d}_1 \geq q^{2(\alpha - 1)}$, implying $\EE_{\mathrm{sp}}^{(2\alpha)} \geq \log q$. \hfill $\square$
\vspace{3pt}

\noindent \hrulefill \textbf{[End of the Proof of Theorem]} 
\vspace{6pt}

We give several remarks regarding Theorem. 
\begin{itemize}
    \item Identities similar to Fig.~\ref{fig:thm1}(a) show that $\TT_{2\alpha}$ has symmetries $V_{g^d}^{(j)} \otimes V_{g^u}^{(j+1)}$ and $V_{h^d}^{(j)} \otimes V_{h^u}^{(j+1)}$ for $1\leq j \leq 2\alpha$, all of which commute. The nontrivial algebra in Eq.~\eqref{alg} arises in $\widetilde{\TT}_{2\alpha}$ due to the partial transpose.

    \item We have $\EE_{\mathrm{sp}}^{(2\alpha)} = \log q$, except for a measure-zero set of mixed states with accidental additional symmetries. 

    \item As discussed in Sec.~\ref{Sec:Thm1_Pre}(c), incoherent Pauli channels $\NN^P$ are non-degenerate for $p \in [0, 1/2)$. Since an injective pure state with nontrivial $G_1 \times G_2$ SPT order remains strongly injective under a brickwork circuit consisting of local non-degenerate channels, the resulting decohered mixed state exhibits spurious TEN for $p \in [0, 1/2)$. 
\end{itemize}

\section{2D Noisy Cluster States on a Square Lattice}

In this Section, we consider the decohered 2D cluster state $\rho_0 = | \psi_0 \rangle \langle \psi_0 |$ on a square lattice. In Sec.~\ref{Sec:2D_Sq_FC}, the fidelity correlator and the stability of mixed-state SSPT order in 2D decohered cluster states are discussed. In Sec.~\ref{Sec:2D_Sq_EN}, the entanglement negativity of 2D decohered cluster states is explained. 

\subsection{Fidelity Correlator}  \label{Sec:2D_Sq_FC}

In Sec.~\ref{Sec:2D_Sq_FC:X}, we exactly compute the fidelity correlator of the $X$-decohered 2D cluster state and demonstrate its exponential decay for $p < 1/2$. In Sec.~\eqref{Sec:2D_Sq_FC:General}, we prove that such exponential decay for $p < 1/2$ appears for general local Pauli noise that preserves subsystem symmetry.

\subsubsection{Fidelity Correlator for \texorpdfstring{$X$}{X}-Noise} \label{Sec:2D_Sq_FC:X}

The fidelity correlator $F_Z^{\text{2D}} (w,h) = F \!\left( \rho_X, \prod_{i\in \square_{wh}} Z_i \rho_X \prod_{i\in \square_{wh}} Z_i \right)$ detects SWSSB of subsystem symmetry, where $\square_{wh}$ represents the four corners of a square of width $w$ and height $h$ located on, say, $B$ sublattice. As explained in the main text, a stat-mech model associated with $F_Z^{\text{2D}} (w,h)$ is the 2D plaquette Ising model (PIM) $H = -\sum_{\square_B} \prod_{i\in \square_B}$, where $\square_B$ denotes elementary plaquettes in the $B$ sublattice. [The other PIM defined on the $A$ sublattice does not contribute to $F_Z^{\text{2D}} (w,h)$, similar to the 1D case.]

On a cylinder of height $L$, subsystem symmetries do not exist along the vertical direction due to the following $ZXZ$ stabilizers (orange shade) at the boundary.
\begin{equation}
    \begin{aligned}
        \includegraphics[height=1.9cm]{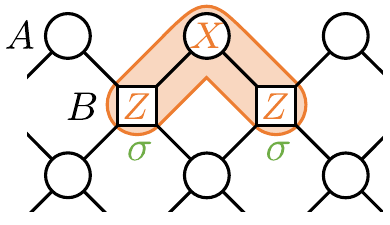}.
    \end{aligned}
\end{equation}
These stabilizers introduce additional boundary Ising interactions $H_\partial$ (green $\sigma\sigma$). Therefore, $F_Z^{\text{2D}} (w,h)$ has a stat-mech expression given by  
\begin{align} \label{2D_Sq_FC_SM}
    F_Z^{\text{2D}} (w,h) = \frac{\sum_S \left[ \langle \sigma_S \rangle_\beta \langle \sigma_{S \triangle \square_{wh}} \rangle_\beta \right]^{1/2}}{\sum_S \langle \sigma_S \rangle_\beta},
\end{align}
where $\beta = -\frac 12 \log (1-2p)$ and $\langle \cdot \rangle_\beta = Z_\beta^{-1} \sum_{\{\sigma\}} (\cdot) e^{\beta \left( \sum_{\square_B} \prod_{i\in \square_B} \sigma_i - H_\partial \right)}$. Here, the summation runs over all subsystem-symmetric subsets $S$ in the $B$ sublattice, i.e., each row and column contains an even number of elements from $S$. Such $S$ can be obtained by taking a modulo-2 union of elementary plaquettes $\square_B$.

Defining new Ising variables as $\tau_{x,1} = \sigma_{x,1}$ and $\tau_{x,y} = \sigma_{x,y-1} \sigma_{x,y}$ for $2\leq y\leq L$ along each column [see Fig.~\ref{fig:2DCluster}(b) of the main text], the plaquette interactions $\prod_{i \in \square} \sigma_i$ map to the Ising interactions $\tau_{x,y} \tau_{x+1,y}$~\cite{mueller2017exact}, whereas the boundary Ising interactions map to interactions non-local along columns, as $\sigma_{x,L} = \prod_{y=1}^L \tau_{x,y}$. However, for $p < 1/2$ (i.e., $\beta < \infty$), these non-local interactions become negligible as we take $L\rightarrow \infty$, which is the limit of interest where the vertical subsystem symmetries are restored. Consequently, the stat-mech model reduces to a stack of independent 1d Ising models (with periodic boundary conditions) in the large $L$ limit. Under this dimensional reduction, subsets $S$ map to $\cup_{y=1}^{L-1} S_y$, where $S_y$ is an even-sized subset of the $y$th 1D Ising chain. This leads to $\langle \sigma_S \rangle_\beta = \prod_{y=1}^{L-1} \langle \tau_{S_y} \rangle_{\beta, \mathrm{Ising}}$, where $\langle \cdot \rangle_{\beta, \mathrm{Ising}}$ denotes the expectation value in the 1D Ising model at inverse temperature $\beta$. As a result, we have
\begin{align}
    F_Z^{\text{2D}} (w,h) \underset{L \rightarrow \infty}{\longrightarrow} \prod_{y = y_1, y_2} \frac{\sum_{S_y} \left[ \langle \tau_{s_y} \rangle_{\beta, \mathrm{Ising}} \langle \tau_{S_y \triangle \{x_1, x_2\}} \rangle_{\beta, \mathrm{Ising}} \right]^{1/2}}{\sum_{S_y} \langle \tau_{S_y} \rangle_{\beta, \mathrm{Ising}}} = [F_Z^{\text{1D}} (h)]^2.
\end{align}
Here, $x_{1,2}$ ($y_{1,2}$) with $|x_1 - x_2| = w$ ($|y_1 - y_2| = h$) is the $x$-coordinates ($y$-coordinates) of the four corners of $\square_{wh}$, and $F_Z^{\text{1D}} (h)$ is the fidelity correlator for the $X$-decohered 1D cluster state [Eq.~\eqref{1D_FC} of the main text] with $|x - y| = h$. Therefore, for $\ < 0.5$, $F_Z^{\text{2D}} (w,h)$ decays exponentially with $h$ in the infinite-cylinder limit. In contrast, at $p = 1/2$, it is easy to show $F_Z^{\text{2D}} (w,h) = 1$.

\subsubsection{Fidelity Correlator for General Pauli Noises}
\label{Sec:2D_Sq_FC:General}

Following steps in Sec.~\ref{Sec:1D_FC:General}, we can analogously show the exponential decay of the fidelity correlator for the 2D cluster state under general local Pauli noises $\NN_j^P [\rho] = (1-p) \rho + p P_j \rho P_j$ that preserves strong subsystem symmetry. This establishes the stability of the mixed-state SSPT order in the decohered 2D cluster state $\rho_P = \prod_{j=1}^{2N} \NN_j^P [\rho_0]$ up to $p = 1/2$. 

Since $[P_j, \prod_{j\in\mathrm{diag}} X_j] = 0$, $P_j$ is a product of finite number of $X$-operators and $\prod_{j\in \square_{A/B}} Z_j$ operators, resulting in a stat-mech model with local interactions $\sigma_{U_j} = \prod_{i\in U_j} \sigma_i$, where $U_i$ is a modulo-2 union of elementary plaquettes $\square_{A/B}$. Also, conjugating $\rho_P$ by charged local Pauli operators $O_i$ (where $i$ are four sites forming a rectangle $\square_{wh}$ of width $w$ and height $h$) is equivalent to conjugating it by $\prod_{i\in R_i} Z_i$, where $R_i$ is some region near site $i$. Therefore, the fidelity correlator $F_O^{\text{2D}} (w,h) = F \!\left( \rho_P, \prod_{i\in \square_{wh}} O_i \rho_P \prod_{i\in \square_{wh}} O_i \right)$ is given by
\begin{align} \label{FC_2D_O}
    F_O^{\text{2D}} (w,h) = \frac{\sum_S \left[ \langle\!\langle \sigma_S \rangle\!\rangle_\beta \langle\!\langle \sigma_{S \triangle \left( \cup_{i\in \square_{wh}} R_i \right)} \rangle\!\rangle_\beta \right]^{1/2}}{\sum_S \langle\!\langle \sigma_S \rangle\!\rangle_\beta},
\end{align}
where $\beta = -\frac 12 \log (1-2p)$ and $\langle\!\langle \cdot \rangle\!\rangle_\beta$ is the expectation value in the stat-mech model $H = -\sum_i \sigma_{U_i}$ at inverse temperature $\beta$, and $S = \bigoplus_i U_i^{e_i}$ ($e_i \in \{0,1\}$) runs over all modulo-2 unions of $U_i$. (As explained in Sec.~\ref{Sec:2D_Sq_FC:X}, For $p < 1/2$, we can safely neglect boundary interactions in the large-height limit.) To ensure that Eq.~\eqref{FC_2D_O} does not vanish trivially for all $p$, the operators $O_{x,y}$ must satisfy $R_x \cup R_y = \bigoplus_i U_i^{r_i}$ for some $r_i \in \{0,1\}$, which we assume below.

Now, introducing new spin variables $\tau_i$ as $\sigma_{U_{i = (x,y)}} = \tau_{x,y} \tau_{x+2,y} \tau_{x,y+2} \tau_{x+2,y+2}$ ($\tau_i \in \{\pm 1\}$) (which can be shown valid using a similar argument in Sec.~\ref{Sec:2D_Sq_FC:X}), we obtain $\langle\!\langle \sigma_S \rangle\!\rangle_\beta = \langle \tau_{S_A} \rangle_\beta \langle \tau_{S_B} \rangle_\beta$, where $S_A$ ($S_B$) are subsets of $j$ in the sublattice $A$ ($B$) with $e_j = 1$, and $\langle \cdot \rangle_\beta$ denotes the expectation value in two copies of the 2D PIM. Similarly, $\langle\!\langle \sigma_{S \triangle \left( \cup_{i\in \square_{wh}} R_i \right)} \rangle\!\rangle_\beta = \langle \tau_{S_A \triangle R_A} \rangle_\beta \langle \tau_{S_B \triangle R_B} \rangle_\beta$, where $R_A$ ($R_B$) are subsets of $j$ in the sublattice $A$ ($B$) with $r_j = 1$. Therefore, Eq.~\eqref{FC_2D_O} factorizes as
\begin{align} \label{FC_2D_O_2}
    F_O^{\text{2D}} (w,h) = \frac{\sum_{S_A} \left[ \langle \tau_{S_A} \rangle_\beta \langle \tau_{S_A \triangle R_A} \rangle_\beta \right]^{1/2}}{\sum_{S_A} \langle \tau_{S_A} \rangle_\beta} \cdot \frac{\sum_{S_B} \left[ \langle \tau_{S_B} \rangle_\beta \langle \tau_{S_B \triangle R_B} \rangle_\beta \right]^{1/2}}{\sum_{S_B} \langle \tau_{S_B} \rangle_\beta},
\end{align}
which is the product of expressions analogous to Eq.~\eqref{2D_Sq_FC_SM}. Now, following the argument in Sec.~\ref{Sec:2D_Sq_FC:X}, Eq.~\eqref{FC_2D_O_2} maps to a product of the fidelity correlators of 1D decohered cluster states [see Eq.~\eqref{FC_O_1D}]. This dimensional reduction shows that $F_O^{\text{2D}} (w,h)$ decays exponentially with $w$ for $p < 1/2$, confirming the stability of mixed-state SSPT order up to the maximal error rate $p = 1/2$.

\subsection{Entanglement Negativity} \label{Sec:2D_Sq_EN}

In Sec.~\ref{Sec:2D_EN:Stab}, we employ stabilizer formalism to compute the entanglement negativity of the $X$-decohered 2D cluster state on a square lattice for $p = 0$ and $1/2$. In Sec.~\ref{Sec:2D_EN:General}, we show that the entanglement negativity of the $X$-decohered 2D cluster state actually reduces to that of the 1D case, confirming $\EE_{\mathrm{sp}} = \log 2$ for $p < 1/2$.

\subsubsection{Stabilizer Formalism for \texorpdfstring{$p = 0$ and $1/2$}{p = 0 and 1/2}}
\label{Sec:2D_EN:Stab}

For $p = 0$, the density matrix $\rho_0$ is a stabilizer state with stabilizers $K_j = X_j \prod_{i\in \partial j} Z_i$. Restricting to the region $R$ above the red dashed line in Fig.~\ref{fig:2DCluster}(c) of the main text, the stabilizers along the boundary become
\begin{equation}
    \begin{aligned}
        \includegraphics[height=2.5cm]{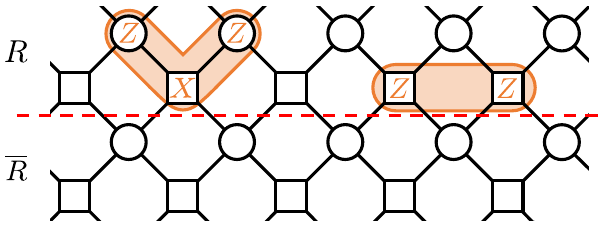}.
    \end{aligned}
\end{equation}
(All stabilizers in the bulk do not contribute to the entanglement negativity as they commute with all restricted stabilizers.) Consequently, the associated matrix $\mathcal{K}_R$ is again given by Eq.~\eqref{KA}, leading to the same entanglement negativity as the 1D cluster state: $\EE_R = N \log 2 - \log 2$, with $\EE_{\text{sp}} = \log 2$. This result aligns with the disentangling argument given in the main text.

For $p = 1/2$, one can easily see that the surviving generators after applying $\prod_{j=1}^{2N} \NN_j^X$ are the subsystem symmetry generators $\prod_{j\in \mathrm{diag}}^N X_j$ along all diagonal lines. From this, we have $\EE_R = 0$.

\subsubsection{Entanglement Negativity for General \texorpdfstring{$p$}{p}} \label{Sec:2D_EN:General}

We establish the presence of spurious TEN in the $X$-decohered 2D cluster state for $p < 1/2$ as follows. The spectrum of $(\rho_X)^{T_R}$ is given by $(\rho_X)^{T_R} \propto \sum_{\{a\}} \!\left[ \prod_j K_j^{a_j} \right]^{T_R} e^{\beta \left(\sum_{\square_A} \prod_{i \in \square_A} \sigma_i + \sum_{\square_B} \prod_{i \in \square_B} \sigma_i \right)}$, where $\beta = -\frac 12 \log (1-2p)$. Since only stabilizers along the entangling surface $\partial R$ are affected by the partial transpose on the upper-half subsystem $R$, we have $\left[ \prod_j K_j^{a_j} \right]^{T_R} = \prod_j K_j^{a_j} (-1)^{\sum_{i\in \partial R} a_i a_{i+1}}$. Introducing new spin variables as 
\begin{align}
    \tau_{x,y} = \begin{cases}
        \sigma_{x,y-1} \sigma_{x,y} & \text{for } y > y_0, \\
        \sigma_{y_0} & \text{for } y = y_0, \\
        \sigma_{x,y} \sigma_{x,y+1} & \text{for } y < y_0
    \end{cases},
\end{align}
where $y_0$ is the $y$-coordinate of an ``anchor'' point $i \in \partial R$, all PIM interactions $\sum_{\square_{A/B}} \prod_{i \in \square_{A/B}} \sigma_i$ transform into horizontal Ising interactions $\tau_{x,y} \tau_{x+1,y}$. This allows the partial-transposed density matrix to factorize as
\begin{align}
    (\rho_X)^{T_R} \underset{L\rightarrow \infty}{\propto} \sum_{\{\tau\}_{y = y_0}}  \prod_x K_j^{a_{x,y_0}} (-1)^{\sum_{i\in \partial R} a_i a_{i+1}} e^{\beta \sum_x \tau_{x,y_0} \tau_{x+1,y_0}} \cdot \sum_{\{\tau\}_{y \neq y_0}} \prod_{x, y\neq y_0} K_{x,y}^{a_{x,y}} e^{\beta \sum_{y\neq y_0} \sum_x \tau_{x,y} \tau_{x+1,y}},
\end{align}
where we have neglected non-local $\tau$ interactions for $p < 1/2$, which become negligible as $L\rightarrow \infty$ (see Sec.~\ref{Sec:2D_Sq_FC:X}). The first factor is nothing but the spectrum of the partial-transposed density matrix of the $X$-decohered 1D cluster state. The second factor is a product of multi-spin correlators in the 1D Ising model and is simply a constant factor that does not contribute to the entanglement negativity. Thus, for $p < 1/2$, the entanglement negativity of the $X$-decohered 2D cluster state reduces to that of the 1D case. Consequently, Theorem and the numerical result shown in Fig.~\ref{fig:1DCluster_SpTEN}(a) of the main text apply to the 2D case as well, yielding $\EE_{\text{sp}} \rightarrow \log 2$ for $p < 1/2$ in the thermodynamic limit. 

\section{Boundary Decoherence in 2D Toric Code}

In Ref.~\cite{lu2024disentangling}, the effect of boundary decoherence on mixed-state long-range entanglement in the toric code is examined. The setup is as follows: consider a 2D toric code on a square lattice with Hamiltonian $H = -\sum_v A_v - \sum_p H_p$, where $A_v = \prod_{i\in \partial v} Z_i$ is the product of $Z$-operators on the four edges adjacent to vertex $v$ (red shade), and $B_p = \prod_{i\in \partial p} X_i$ is the product of $X$-operators on the four edges enclosing plaquette $p$ (blue shade). The system is bipartitioned along the yellow dashed line below, with decoherence applied only to the qubits along the bipartition boundary (yellow circles) via a local quantum channel. 
\vspace{-5pt}
\begin{equation}
    \begin{aligned}
        \includegraphics[height=4.7cm]{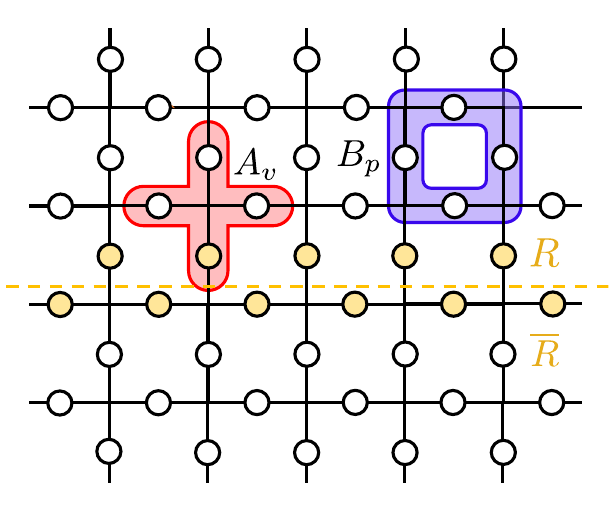}.
    \end{aligned}
\end{equation}

\vspace{-5pt} Under both $X$-noise and $Z$-noise with respective rates $p_x$ and $p_z$ at the boundary, it is shown in Ref.~\cite{lu2024disentangling} that the entanglement negativity is determined by the boundary portion of the partial-transposed density matrix:
\begin{align} \label{rho_partial}
    \rho_\partial^{T_R} \propto \sum_{\{a\}} \sum_{\{b\}} \prod_{j=1}^N A_j^{a_j} \prod_{j=1}^N B_j^{b_j} (-1)^{\sum_{i=1}^N a_i (b_{i-1} + b_{i+1})} e^{\beta_x \sum_{i=1}^N \sigma_i \sigma_{i+1}} e^{\beta_z \sum_{i=1}^N \tau_i \tau_{i+1}},
\end{align}
where $\beta_{x,z} = -\frac 12 \log(1 - 2p_{x,z})$ and $2N$ is the length of the bipartition boundary. Here, $a_j, b_j \in \{0,1\}$ indicates the presence or absence of the stabilizers $A_j$ and $B_j$ along the boundary, with the corresponding Ising spin variables defined as $\sigma_j = 1 - 2a_j$ and $\tau_j = 1 - 2b_j$. Ref.~\cite{lu2024disentangling} has left an open question of the critical error threshold at which mixed-state long-range entanglement between $R$ and its complement $\overline{R}$ breaks down.

As a by-product of our study, we resolve this question. Note that the spectrum of Eq.~\eqref{rho_partial} is equal to that of the partial-transposed density matrix for an $X$-decohered 1D cluster state, where qubits in sublattice $A$ ($B$) decohere at rate $p_x$ ($p_z$). Consequently, in this setup,  the TEN of the decohered 2D toric code coincides with the spurious TEN of the $X$-decohered 1D cluster state. When $p_x = p_z \equiv p$, Eq.~\eqref{rho_partial} reduces to Eq.~\eqref{rhoDX_TA} in the main text. Thus, Fig.~\ref{fig:1DCluster_SpTEN}(a) provides numerical evidence that mixed-state long-range entanglement between $R$ and $\overline{R}$ in the 2D toric code remains robust up $p = 1/2$. For general cases with $p_x \neq p_z$, Theorem implies that the same conclusion holds.

\end{document}